\documentclass[notitlepage]{revtex4} 
\usepackage{graphicx}

\usepackage{hyperref} 					 
\usepackage{listings} 					 
\usepackage{color}    					 
\usepackage{natbib}   					 
\usepackage{subfig} 					 

\usepackage{booktabs}
\usepackage[ansinew]{inputenc}				     
\usepackage{psfrag}
\usepackage{pstricks}
\usepackage{amsfonts}
\usepackage{amsmath}
\usepackage{bm}
\usepackage{lipsum}
\usepackage{pgf}
\usepackage{tikz}
\usetikzlibrary{patterns}

\usepackage{fancyvrb}
\usepackage{listings}
\usepackage{color} 
\definecolor{mygreen}{RGB}{28,172,0} 
\definecolor{mylilas}{RGB}{170,55,241}
\usepackage{url} 
\usepackage[normalem]{ulem} 
\useunder{\uline}{\ul}{} 

\usepackage{array}
\newcolumntype{C}[1]{>{\centering\arraybackslash}m{#1}}

\usepackage{algpseudocode}
\usepackage{color}

\usepackage{filecontents}

\begin{document}

\title{Compact 200 line MATLAB code for inverse design in photonics by topology optimization: tutorial}

\author{Rasmus E. Christiansen,$^{1,2}$}
\email{raelch@mek.dtu.dk}

\author{Ole Sigmund,$^{1,2}$}

\affiliation{$^{1}$ Department of Mechanical Engineering, Technical University of Denmark, Nils Koppels All\'{e}, Building 404, 2800 Kongens Lyngby, Denmark}
\affiliation{$^{2}$ NanoPhoton---Center for Nanophotonics, Technical University of Denmark, {\O}rsteds Plads 345A, DK-2800 Kgs. Lyngby, Denmark.}

\begin{abstract}
We provide a compact 200 line MATLAB code demonstrating how topology optimization (TopOpt) as an inverse
design tool may be used in photonics, targeting the design of two-dimensional dielectric metalenses and a metallic
reflector as examples. The physics model is solved using the finite element method, and the code utilizes MATLAB’s
fmincon algorithm to solve the optimization problem. In addition to presenting the code itself, we briefly discuss
a number of extensions and provide the code required to implement some of these. Finally, we demonstrate the
superiority of using a gradient-based method compared to a genetic-algorithm-based method (using MATLAB’s ga
algorithm) for solving inverse design problems in photonics. The MATLAB software is freely available in the paper
and may be downloaded from \url{https://www.topopt.mek.dtu.dk}.
\end{abstract}

\maketitle

\section{Introduction}

This paper details a 200 line MATLAB code, which demonstrates how density-based Topology Optimization (TopOpt) can be applied to photonics design. The code is written for scientists and students with a basic knowledge of programming, numerical modelling and photonics, who desire to start using inverse design in their research. We briefly detail the model of the physics, followed by the discretized TopOpt design problem (Sec.~\ref{SEC:PHYSICS_OPTIMIZATION}). The MATLAB code is then explained in detail (Sec.~\ref{SEC:MATLAB_CODE}), followed by two application examples providing the reader with targets for reproduction (Sec.~\ref{SEC:DESIGN_EXAMPLES}). Then, a number of possible extensions are discussed and code-snippets for easy implementation are provided \textcolor{black}{along with design examples} (Sec.~\ref{SEC:ADVANCED}). \textcolor{black}{Finally, we demonstrate the superiority of using gradient-based TopOpt compared to a genetic algorithm when solving a photonic design problem (Sec.~\ref{SEC:GA_VS_TOPOPT}).}  

Topology optimization \cite{BOOK_TOPOPT_BENDSOE} as an inverse design tool was first developed in the context of solid mechanics in the late 1980s \cite{BENDSOE_KIKUCHI_1988}. Since its inception the method has developed rapidly and expanded across most areas of physics \cite{ALEXANDERSEN_ANDREASEN_2020,DILGEN_ET_AL_2019,LUNDGAARD_2019,JENSEN_SIGMUND_2011}. Over the last two decades the interest in applying TopOpt for photonics has increased rapidly \cite{MOLESKY_2018} with applications within cavity design \cite{LIANG_JOHNSON_2013,WANG_2018}, photonic demultiplexers \cite{PIGGOTT_2015}, metasurfaces \cite{ZIN_2019,CHUNG_MILLER_2020} and topological insulators \cite{CHRISTIANSEN_NP_2019} to name a few. While the interest in TopOpt within the photonics community has grown markedly \textcolor{black}{in recent years}, significant barriers hinder newcomers to the field from adopting the tool in their work. These are: the required knowledge of numerical modelling, of advanced mathematical concepts and scientific programming experience. This paper seeks to lower these barriers by providing the reader with a simple 2D finite element based MATLAB implementation of TopOpt for photonics, which is straightforwardly extendable to a range of other design problems. Within the field of structural optimization in mechanics similar simple MATLAB codes \cite{SIGMUND_2001,ANDREASSEN_2011,FERRARI_2020} have proven themselves highly successful in raising the awareness of TopOpt and serving as a basic platform for further expansion of the method, thus broadening its application as a design tool and successfully driving the field forward. 

For readers who are less interested in the programming and method development aspects of TopOpt \textcolor{black}{as an inverse design tool}, we have authored a parallel tutorial paper on TopOpt for photonics applications, utilizing the GUI based commercial finite element software COMSOL Multiphysics as the numerical tool to model the physics and solve the optimization problem~\cite{CHRISTIANSEN_SIGMUND_COMSOL_2020}. 

\section{The Physics and the Discretized Optimization Problem} \label{SEC:PHYSICS_OPTIMIZATION}

\begin{figure}[h!]
	\centering
	{
		\includegraphics[width=0.7\textwidth]{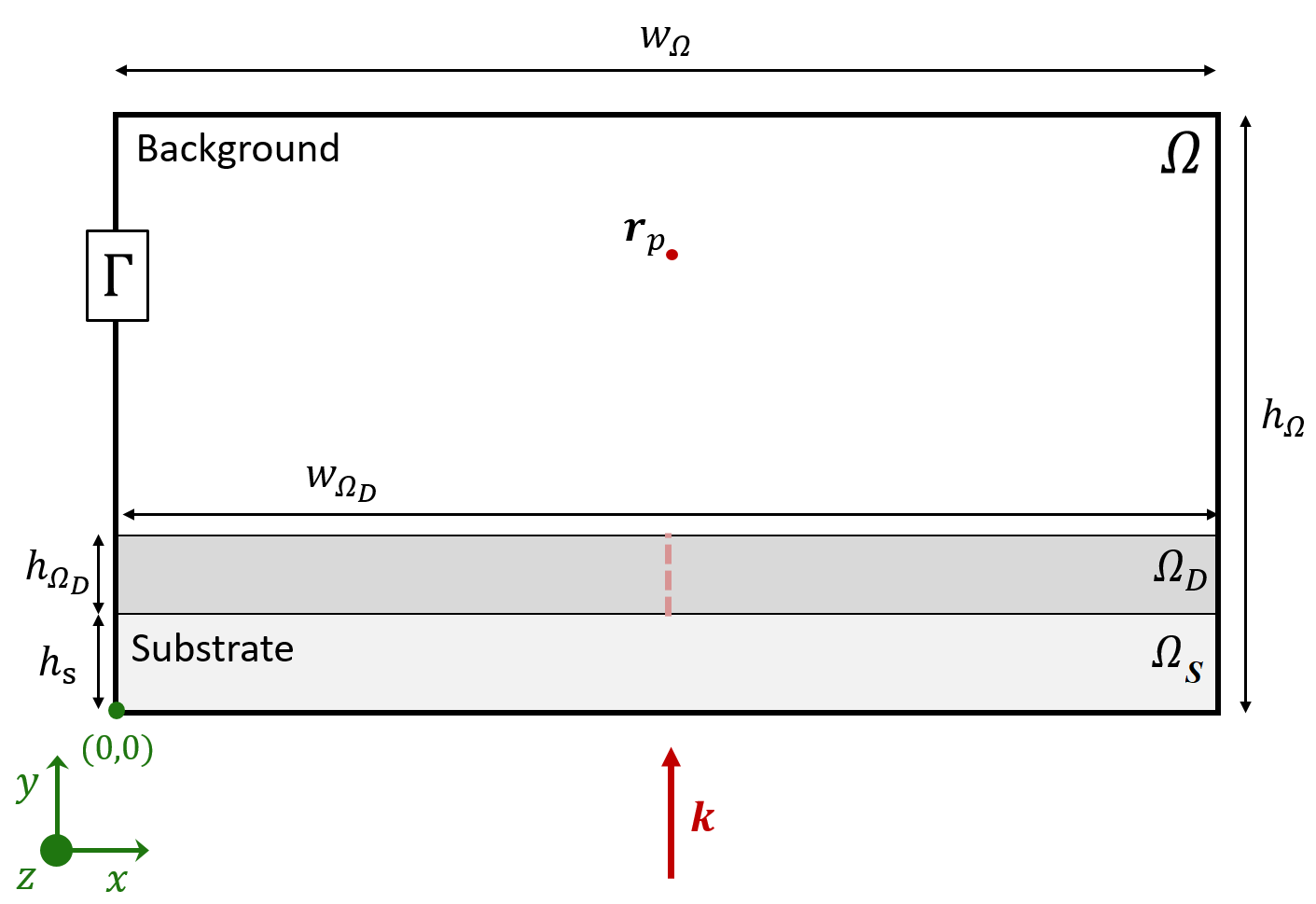} \caption{Model domain, $\Omega$, of height $h_\Omega$ and width $w_\Omega$ with a designable region, $\Omega_D$, of height $h_{\Omega_D}$ and width $w_{\Omega_D}$ on top of a substrate, $\Omega_S$, of height $h_s$. \label{FIG:MATLAB_META_LENS_MODEL_PROBLEM}}
	}
\end{figure}

We model the physics in the rectangular domain $\Omega$, with the boundary $\Gamma$, (see Fig.~\ref{FIG:MATLAB_META_LENS_MODEL_PROBLEM}) using Maxwell's equations, assuming time-harmonic temporal behaviour. \textcolor{black}{We define a subset of $\Omega$ as the design domain and denote this region $\Omega_D$.} We assume out-of-plane ($z$-direction) material invariance and that all involved materials are linear, static, homogeneous, isotropic, non-dispersive, non-magnetic and without inherent polarization. Finally, we assume out-of-plane polarization of the electric field (TE polarization). From these assumptions we derive a two-dimensional Helmholtz-type partial differential equation for the out-of-plane component of the electric field \textcolor{black}{in $\Omega$}, \vspace{-5pt}

\begin{eqnarray} \label{EQN:FORWARD_PROBLEM}
\nabla \cdot \left( \nabla E_z(\textbf{r}) \right) + k^2 \varepsilon_r(\textbf{r}) E_z(\textbf{r}) = F(\textbf{r}), \ \ \textbf{r} \in \Omega \in \mathbb{R}^2,
\end{eqnarray}

\noindent where $E_z$ denotes the z-component of the electric field, $k = \frac{2\pi}{\lambda}$ is the wavenumber with $\lambda (=$ \verb|lambda| in the \verb|top200EM| interface$)$ being the wavelength, $\varepsilon_r$ denotes the relative electric permittivity and $F$ denotes a forcing term used to introduce an incident plane wave from the bottom boundary  \textcolor{black}{of $\Omega$}. We apply first order absorbing boundary conditions on all four exterior boundaries,  \vspace{-5pt}

\begin{eqnarray}
\textbf{n} \cdot \nabla E_z(\textbf{r}) = -\mathrm{i} k E_z(\textbf{r}), \ \ \textbf{r} \in \Gamma,
\end{eqnarray}

\noindent where $\textbf{n}$ denotes the surface normal and $\mathrm{i}$ the imaginary unit. Note that first order boundary conditions are not as accurate as \textcolor{black}{certain} other boundary conditions, e.g. perfectly matched layers \cite{BERENGER_1994}, however they are \textcolor{black}{conceptually} simpler \textcolor{black}{and simpler to implement}. Next, we introduce a design field $\xi(\textbf{r}) \in [0,1]$ to control the material distribution in \textcolor{black}{$\Omega$} through the interpolation \textcolor{black}{function}, \vspace{-5pt}

\begin{eqnarray}
\varepsilon_r(\xi(\textbf{r})) =  1 + \xi(\textbf{r}) \left( \varepsilon_{r,m} - 1 \right) - \mathrm{i} \ \alpha \ \xi(\textbf{r}) \left( 1 - \xi(\textbf{r}) \right), \ \ \textcolor{black}{ \textbf{r} \in \Omega},
\end{eqnarray}

\noindent where \textcolor{black}{it is assumed that the background has the value $\varepsilon_r = 1$ (e.g. air)} and where $\varepsilon_{r,m} (=$ \verb|eps_r|$)$ denotes the relative permittivity of the material used for the structure under design and $\alpha$ is a problem dependent scaling factor. The non-physical imaginary term discourages intermediate values of $\xi$ in the design for the \textcolor{black}{focussing} problem at hand, \textcolor{black}{by introducing attenuation} \cite{JENSEN_SIGMUND_2005}. 

\textcolor{black}{As the baseline example, we consider the design of a monochromatic focussing metalens situated in $\Omega_D$}. To this end, we select the magnitude of $\vert E_z \vert^2$ at a point in space $\textbf{r}_p (=$ \verb|targetXY|$)$ as the figure of merit (FOM) \textcolor{black}{denoted} $\Phi$, i.e. \vspace{-5pt}

\begin{eqnarray}
\Phi(\xi(\textbf{r}),\textbf{r}_p) = \vert E_z(\xi(\textbf{r}),\textbf{r}_p) \vert^2 = E_z(\xi(\textbf{r}),\textbf{r}_p)^{*} E_z(\xi(\textbf{r}),\textbf{r}_p),
\end{eqnarray}

\noindent where $\bullet^{*}$ denotes the complex conjugate. 

The model equation, boundary conditions, material-interpolation function and figure of merit are \textcolor{black}{all} discretized using the finite element method (FEM) \cite{BOOM_FEM_JIN} using $\mathcal{N}_e (= $\verb|nElX| $\cdot$ \verb|nElY|$)$ bi-linear quadratic elements. \textcolor{black}{The discretized model uses} nodal degrees of freedom (DOFs) for $E_z$ \textcolor{black}{and $F$ as well as} elementwise constant DOFs for $\xi$, \textcolor{black}{with $\mathcal{N}_D$ of the elements situated in $\Omega_D$}. The following constrained continuous optimization problem is formulated for the discretized problem, \vspace{-5pt}

\begin{eqnarray} 
&\underset{\boldmath{\xi}}{\max} \textcolor{white}{ : } \ \ & \boldmath{\Phi} = \textcolor{black}{\textbf{E}_z^{\dagger} \textbf{P} \textbf{E}_z}, \label{EQN:OPTIMIZATION_PROBLEM} \\
&\mathrm{s.t. :} & \textbf{S}(\boldmath{\varepsilon}_r) \textbf{E}_z =  \left(\sum_{e=1}^{\mathcal{N}_e} \textbf{S}_e(\varepsilon_{r,e})\right) \textbf{E}_z = \textbf{F}, \nonumber \\
&\mathrm{\textcolor{white}{s.t.} :} &\boldmath{\varepsilon}_{r,j}=  1 + \bar{\tilde{\boldmath{\xi}}}_j \left( \varepsilon_{r,m} - 1 \right) - \mathrm{i} \ \bar{\tilde{\boldmath{\xi}}}_j \left( 1 -  \bar{\tilde{\boldmath{\xi}}}_j \right) \ \ \forall \ j \in \textcolor{black}{\lbrace 1,2,...,\mathcal{N}_e\rbrace}, \nonumber \\
&\mathrm{\textcolor{white}{s.t.} :} & 0<\boldmath{\xi}_j <1 \ \ \forall \ j \in \textcolor{black}{\lbrace 1,2,...,\mathcal{N}_D\rbrace}, \nonumber \\
&\mathrm{\textcolor{white}{s.t.} :} & \textcolor{black}{\xi = 0 \ \ \forall \ \ \textbf{r} \in \Omega / \lbrace \Omega_D, \Omega_S \rbrace \ \ \vee \ \  \xi = 1 \ \ \forall \ \ \textbf{r} \in \Omega_S,} \nonumber
\end{eqnarray}

\noindent where $\textbf{E}_z$ and $\textbf{F}$ are vectors containing the nodal DOFs for the electric field and forcing term and $\boldmath{\xi}(=$ \verb|dVs|$)$ and $\bar{\tilde{\boldmath{\xi}}}$ are vectors of element DOFs for the design field and the physical filtered and thresholded field (eqs.~(\ref{EQN:FILTER_CONVOLUTION})-(\ref{EQN:THRESHOLD_FUNCTION})), respectively. 

\textcolor{black}{The diagonal selection matrix $\textbf{P}$ weights the $\textbf{E}_z$-DOFs that enter} $\boldmath{\Phi} (=$ \verb|FOM|$)$. \textcolor{black}{In the baseline example it has four $\frac{1}{4}$ entries for the selected element's nodal DOFs. I.e.} it selects the field intensity at the focal point, which for simplicity is taken to be at the center of a single finite element. Finally, $\bullet^{\dagger}$ denotes the conjugate transpose.


To  \textcolor{black}{ameliorate} numerical issues, such as pixel-by-pixel design variations, and to introduce a weak sense of geometric length scale, a filter and threshold scheme is applied to $\boldmath{\xi}$, before using it to interpolate the material parameters \cite{GUEST_ET_AL_2004,WANG_ET_AL_2011,CHRISTIANSEN_SMO_2015}. First, the following convolution-based filter operation is applied, \vspace{-5pt}

\begin{eqnarray}
\tilde{\boldmath{\xi}}_h = \frac{\underset{k \in \mathcal{B}_{e,h}}{\sum} w(\textbf{r}_h - \textbf{r}_k)A_k \boldmath{\xi}_k}{\underset{k \in \mathcal{B}_{e,h}}{\sum} w(\textbf{r}_h - \textbf{r}_k)A_k}, \ \ w(\textbf{r}) =  \begin{cases}
r_f - \vert \textbf{r} \vert \ \ \ \ \forall \ \vert \textbf{r} \vert \leq r_f \\
0
\end{cases}, \ \ r_f  \geq 0, \ \ \textbf{r} \in \Omega. \label{EQN:FILTER_CONVOLUTION}
\end{eqnarray}

\noindent Here $A_k$ denotes the area of the $k$'th element and $r_f (=$ \verb|fR|$)$ the desired spatial filtering radius, finally $\mathcal{B}_{e,h}$ denotes the $h$'th set of finite elements who's center point is within $r_f$ of the $h$'th element. Second, a smoothed approximation of the Heaviside function is applied to the \textcolor{black}{filtered} design variables as, \vspace{-5pt}

\begin{eqnarray}
\bar{\tilde{\boldmath{\xi}}}_h = H({\tilde{\boldmath{\xi}}}_h) = \frac{\tanh(\beta \cdot \eta) + \tanh(\beta \cdot (\tilde{\boldmath{\xi}}_h - \eta))}{\tanh(\beta \cdot \eta) + \tanh(\beta \cdot (1 - \eta))}, \ \ \beta \in [1,\infty[, \ \ \eta \in [0,1].  \label{EQN:THRESHOLD_FUNCTION}
\end{eqnarray}

\noindent Here $\beta$ and $\eta$ control the threshold sharpness and value, respectively. \\

\noindent Adjoint sensitivity analysis \cite{TORTORELLI_ET_AL_1994,JENSEN_SIGMUND_2011} is carried out to compute the design sensitivities, utilizing the chain rule for the filter and threshold steps as \cite{Duhring_2008,CHRISTIANSEN_SMO_2015},\vspace{-5pt}

\begin{eqnarray} \label{EQN:SENSITIVITES}
\frac{\mathrm{d} \boldmath{\Phi}}{\mathrm{d} \boldmath{\xi}_h} = \underset{k\in \mathcal{B}_{e,h} }{\sum} \frac{\partial \tilde{\boldmath{\xi}}_k}{\partial \boldmath{\xi}_h} \frac{\partial \bar{\tilde{\boldmath{\xi}}}_k}{\partial \tilde{\boldmath{\xi}}_k} \frac{\mathrm{d} \boldmath{\Phi}}{\mathrm{d} \bar{\tilde{\boldmath{\xi}}}_k}, \ \ \ \ \ \frac{\mathrm{d} \boldmath{\Phi}}{\mathrm{d} \bar{\tilde{\boldmath{\xi}}}_k} = 2 \ \Re\left(\boldmath{\lambda}^{\mathrm{T}} \frac{\partial \textbf{S}}{\partial \bar{\tilde{\boldmath{\xi}}}_k} \textbf{E}_z \right),
\end{eqnarray}

\noindent where $\Re$ denotes the real part, \textcolor{black}{$\bullet^T$} the transpose and $\boldmath{\lambda}$ a vector obtained by solving, \vspace{-5pt}

\begin{eqnarray} \label{EQN:ADJOINT_LINEAR_SYSTEM}
\textbf{S}^{\mathrm{T}}\boldmath{\lambda} = -\frac{1}{2}\left( \frac{\partial \boldmath{\Phi}}{\partial \textbf{E}_{z,\Re}} - \mathrm{i} \frac{\partial \boldmath{\Phi}}{\partial \textbf{E}_{z,\Im}} \right)^{\mathrm{T}} \ \mathrm{ with } \ \ \textbf{E}_z = \textbf{E}_{z,\Re} + \mathrm{i} \ \textbf{E}_{z,\Im},
\end{eqnarray}

\noindent where $\Im$ denotes the imaginary part. The $m$'th entry of the right-hand side in eq.~(\ref{EQN:ADJOINT_LINEAR_SYSTEM}) is given by, \vspace{-5pt}

\begin{eqnarray} \label{EQN:ADJOINT_RHS}
\left( \frac{\partial \boldmath{\Phi}}{\partial \textbf{E}_{z,\Re}} - \mathrm{i} \frac{\partial \boldmath{\Phi}}{\partial \textbf{E}_{z,\Im}} \right)_m^{\mathrm{T}} = \textbf{P}_{m,m} \left( 2 {(\textbf{E}_{z,\Re})}_m - 2 \mathrm{i} \ {(\textbf{E}_{z,\Im})}_m \right). 
\end{eqnarray}

\noindent \textcolor{black}{The derivations of the expression for $\frac{\mathrm{d} \boldmath{\Phi}}{\mathrm{d} \bar{\tilde{\boldmath{\xi}}}_k}$ in eq.~(\ref{EQN:SENSITIVITES}) and for eq.~(\ref{EQN:ADJOINT_LINEAR_SYSTEM}) are given in Appendix~\ref{APN:ADJOINT_SENSITIVITY_ANALYSIS}.}\\

\textcolor{black}{The fundamental advantage of using the adjoint approach to compute the sensitivities of the FOM with respect to the design variables is that only one single additional system of equations, (namely eq.~(\ref{EQN:ADJOINT_LINEAR_SYSTEM})), must be solved. Hence, the sensitivity information is obtained at approximately the same computational cost as the one required to compute the field information itself. In fact, for the examples treated here, it is possible to reuse the LU-factorization used to compute the field information (second line of eq.~(\ref{EQN:OPTIMIZATION_PROBLEM})), making the computational cost associated with computing the sensitivity of the FOM almost ignorable.}

\section{The MATLAB Code} \label{SEC:MATLAB_CODE}

The design problem stated in eq.~(\ref{EQN:OPTIMIZATION_PROBLEM}) is implemented in \verb|top200EM| (see the full code in Appendix~\ref{APN:MATLAB_CODE}), \textcolor{black}{which has the interface}: \\

\begin{verbatim}
function [dVs,FOM] = ... 
top200EM(targetXY,dVElmIdx,nElX,nElY,dVini,eps_r,lambda,fR,maxItr);
\end{verbatim}

\noindent The function takes the input parameters:
	\begin{itemize}
		\item[] \verb|targetXY|: \textcolor{black}{two-values 1D-array} with the x- and y-position of the finite element containing the focal point.
		\item[] \verb|dVElmIdx|: 1D-array of indices of the finite elements which are designable, i.e. $\Omega_D$. 
		\item[] \verb|nElX|: Number of finite elements in the x-direction.
		\item[] \verb|nElY|: Number of finite elements in the y-direction.
		\item[] \verb|dVini|: Initial guess for \textcolor{black}{discretized} $\boldmath{\xi}$-field. Accepts a scalar for all elements or a 1D-array of identical length to \verb|dVElmIdx|.
		\item[] \verb|eps_r|: Relative permittivity for the material constituting the structure under design.
		\item[] \verb|lambda|: The targeted wavelength, $\lambda$, measured in \textcolor{black}{number of} finite elements.
		\item[] \verb|fR|: Filter radius, $r_f$, measured in \textcolor{black}{number of} finite elements.
		\item[] \verb|maxItr|: Maximum number of iterations allowed by \verb|fmincon| for solving the optimization problem in eq.~(\ref{EQN:OPTIMIZATION_PROBLEM}).
	\end{itemize}
\noindent And returns the output parameters:
	\begin{itemize}
		\item[] \verb|dVs|: \textcolor{black}{1D-array of the optimized discretized $\boldmath{\xi}$-field} in the design domain, $\Omega_D$. 
		\item[] \verb|FOM|: \textcolor{black}{Value of the} figure of merit. 
	\end{itemize}

\noindent During execution, the data related to the physics, the discretization and the filter and threshold operations are stored in the structures \verb|phy|, \verb|dis| and \verb|filThr|, respectively. \\

\noindent The spatial scaling, threshold sharpness $\beta$ and threshold level $\eta$ are hard coded in \verb|top200EM| as,

\begin{itemize}
	\item[] 
\begin{verbatim}
% SETUP OF PHYSICS PARAMETERS
phy.scale = 1e-9; % Scaling finite element side length to nanometers

% SETUP FILTER AND THRESHOLDING PARAMETERS
filThr.beta = 5; % Thresholding sharpness
filThr.eta = 0.5; % Thresholding level
\end{verbatim}
\end{itemize}

\noindent \textbf{Note:} For simplicity the code uses the unit of nanometers to measure length and the finite elements are taken to have a side length of 1 nm. This may be changed by changing the scaling parameter \verb|phy.scale|. \\

\noindent The algorithm used to solve the design problem is MATLABs \verb|fmincon|,

\begin{itemize}
	\item[] 
\begin{verbatim}
[dVs,~] = fmincon(FOM,dVs(:),[],[],[],[],LBdVs,UBdVs,[],options);
\end{verbatim}
\end{itemize}

\noindent with the design-variable bounds and options set up as,

\begin{itemize}
	\item[] 
\begin{verbatim}
LBdVs = zeros(length(dVs),1); % Lower bound on design variables
UBdVs = ones(length(dVs),1); % Upper bound on design variables
options = optimoptions('fmincon','Algorithm','interior-point',...
'SpecifyObjectiveGradient',true,'HessianApproximation','lbfgs',...
'Display','off','MaxIterations',maxItr,'MaxFunctionEvaluations',maxItr);
\end{verbatim}
\end{itemize}

\noindent The FOM and sensitivites provided to \verb|fmincon| are computed using the inline function:

\begin{itemize}
	\item[] 
\begin{verbatim}
FOM = @(dVs)OBJECTIVE_GRAD(dVs,dis,phy,filThr);
\end{verbatim}
\end{itemize}

\noindent The discretized design field is distributed in the model domain with a hard coded background of air in the top 90\% of the domain and solid material in the bottom 10\% as,

\begin{itemize}
	\item[] 
\begin{verbatim}
% DISTRIBUTE MATERIAL IN MODEL DOMAIN BASED ON DESIGN FIELD
dFP(1:dis.nElY,1:dis.nElX) = 0; % Design field in physics, 0: air
dFP(dis.nElY:-1:ceil(dis.nElY*9/10),1:dis.nElX) = 1; % 1: material
dFP(dis.dVElmIdx(:)) = dVs; % Design variables inserted in design field
\end{verbatim}
\end{itemize}

\noindent Followed by the application of the filter and threshold operations and the material interpolation,

\begin{itemize}
	\item[] 
\begin{verbatim}
% COMPUTE MATERIAL FIELD FROM DESIGN FIELD
dFPS = DENSITY_FILTER(filThr.filKer,filThr.filSca,dFP,ones(dis.nElY,dis.nElX));
dFPST = THRESHOLD( dFPS, filThr.beta, filThr.eta);
[A,dAdx] = MATERIAL_INTERPOLATION(phy.eps_r,dFPST,1.0); % Material field
\end{verbatim}
\end{itemize}

\noindent The system matrix for the state equation is constructed,

\begin{itemize}
	\item[] 
\begin{verbatim}
% CONSTRUCT SYSTEM MATRIX
[dis,F] = BOUNDARY_CONDITIONS_RHS(phy.k,dis,phy.scale);
dis.vS = reshape(dis.LEM(:)-phy.k^2*dis.MEM(:)*(A(:).'),16*dis.nElX*dis.nElY,1);
SysMat = sparse([dis.iS(:);dis.iBC(:)],[dis.jS(:);dis.jBC(:)],[dis.vS(:);dis.vBC(:)]);
\end{verbatim}
\end{itemize}

\noindent The state system is solved using LU-factorization,

\begin{itemize}
	\item[] 
\begin{verbatim}
% SOLVING STATE SYSTEM: SysMat * Ez = F
[L,U,Q1,Q2] = lu(SysMat); % LU - factorization
Ez = Q2 * (U\(L\(Q1 * F)));  Ez = full(Ez); % Solving
\end{verbatim}
\end{itemize}

\noindent The FOM is computed,

\begin{itemize}
	\item[] 
\begin{verbatim}
% FIGURE OF MERIT 
P = sparse(dis.edofMat(dis.tElmIdx,:),dis.edofMat(dis.tElmIdx,:),1/4,...
(dis.nElX+1)*(dis.nElY+1),(dis.nElX+1)*(dis.nElY+1)); % Weighting matrix
FOM = Ez' * P * Ez; % Solution in target element
\end{verbatim}
\end{itemize}

\noindent The adjoint system is solved by reusing the LU-factorization from the state problem,

\begin{itemize}
	\item[] 
\begin{verbatim}
% ADJOINT RIGHT HAND SIDE (0th-order quadrature)
AdjRHS = P*(2*real(Ez) - 1i*2*imag(Ez));

% SOLVING THE ADJOING SYSTEM: S.' * AdjLambda = AdjRHS
AdjLambda = (Q1.') * ((L.')\((U.')\((Q2.') * (-1/2*AdjRHS)))); % Solving
\end{verbatim}
\end{itemize}

\noindent The sensitivities in $\Omega$ are computed and filtered, and the values in $\Omega_D$ extracted,

\begin{itemize}
	\item[] 
\begin{verbatim}
% SENSITIVITIES
dis.vDS = reshape(-phy.k^2*dis.MEM(:)*(dAdx(:).'),16*dis.nElX*dis.nElY,1); 
DSdx = sparse(dis.iElFull,dis.jElFull,dis.vDS); % Constructing dS/dx
DSdxMulV = DSdx * Ez(dis.idxDSdx); % Computing dS/dx * Field values
DsdxMulV = sparse(dis.iElSens,dis.jElSens,DSdxMulV);
sens = 2*real(AdjF(dis.idxDSdx).' * DsdxMulV); % Computing sensitivites
sens = full(reshape(sens,dis.nElY,dis.nElX));

% FILTERING SENSITIVITIES
DdFSTDFS = DERIVATIVE_OF_THRESHOLD( dFPS, filThr.beta, filThr.eta);
sensFOM = DENSITY_FILTER(filThr.filKer,filThr.filSca,sens,DdFSTDFS);

% EXTRACTING SENSITIVITIES FOR DESIGN REGION
sensFOM = sensFOM(dis.dVElmIdx);
\end{verbatim}
\end{itemize}

\noindent Finally, $E_z(\textbf{r})$ and $\xi(\textbf{r})$ are plotted and the current FOM-value printed, 

\begin{itemize}
	\item[] 
\begin{verbatim}
% PLOTTING AND PRINTING
figure(1); % Field intensity, |Ez|^2
imagesc((reshape(Ez.*conj(Ez),dis.nElY+1,dis.nElX+1))); colorbar; axis equal;
figure(2); % Physical design field
imagesc(1-dFPST); colormap(gray); axis equal; drawnow;
disp(['FOM: ' num2str(-FOM)]); % Display FOM value
\end{verbatim}
\end{itemize}

\noindent After the TopOpt procedure is finished, a thresholded version of the final design is evaluated and the resulting $\vert E_z \vert^2$ field and design field are plotted.

\begin{itemize}
	\item[] 
\begin{verbatim}
% FINAL BINARIZED DESIGN EVALUATION
filThr.beta = 1000; 
disp('Black/white design evaluation:')
[obj_2,dFPST_2,F_2] = OBJECTIVE_GRAD(DVini(:),dis,phy,filThr); 
\end{verbatim}
\end{itemize}

\section{Using the Code} \label{SEC:DESIGN_EXAMPLES}

Next, we demonstrate how to use \verb|top200EM| by designing a focusing metalens as follows. 

\subsection{Designing a Metalens} \label{SEC:DESIGN_METALENS}

\noindent First, we define the domain size \textcolor{black}{in terms of the number of finite elements in each spatial direction} and the element indices for the design domain. 

\begin{itemize}
	\item[] 
\begin{verbatim}
% DESIGN FIELD INDICES
DomainElementsX = 400; 
DomainElementsY = 200; 
DesignThicknessElements = 15;
DDIdx = repmat([1:DomainElementsY:DomainElementsX*DomainElementsY],...
DesignThicknessElements,1);
DDIdx = DDIdx+repmat([165:165+DesignThicknessElements-1]',1,DomainElementsX);
\end{verbatim}
\end{itemize}

\noindent Second, the optimization problem is solved by executing the command,

\begin{itemize}
	\item[] 
\begin{verbatim}
[DVs,obj]=top200EM([200,80],DDIdx,DomainElementsX,DomainElementsY,...
                   0.5,3.0,35,6.0,200);
\end{verbatim}
\end{itemize}

\begin{figure}[h!]
	\centering
	{
		\includegraphics[width=0.85\textwidth]{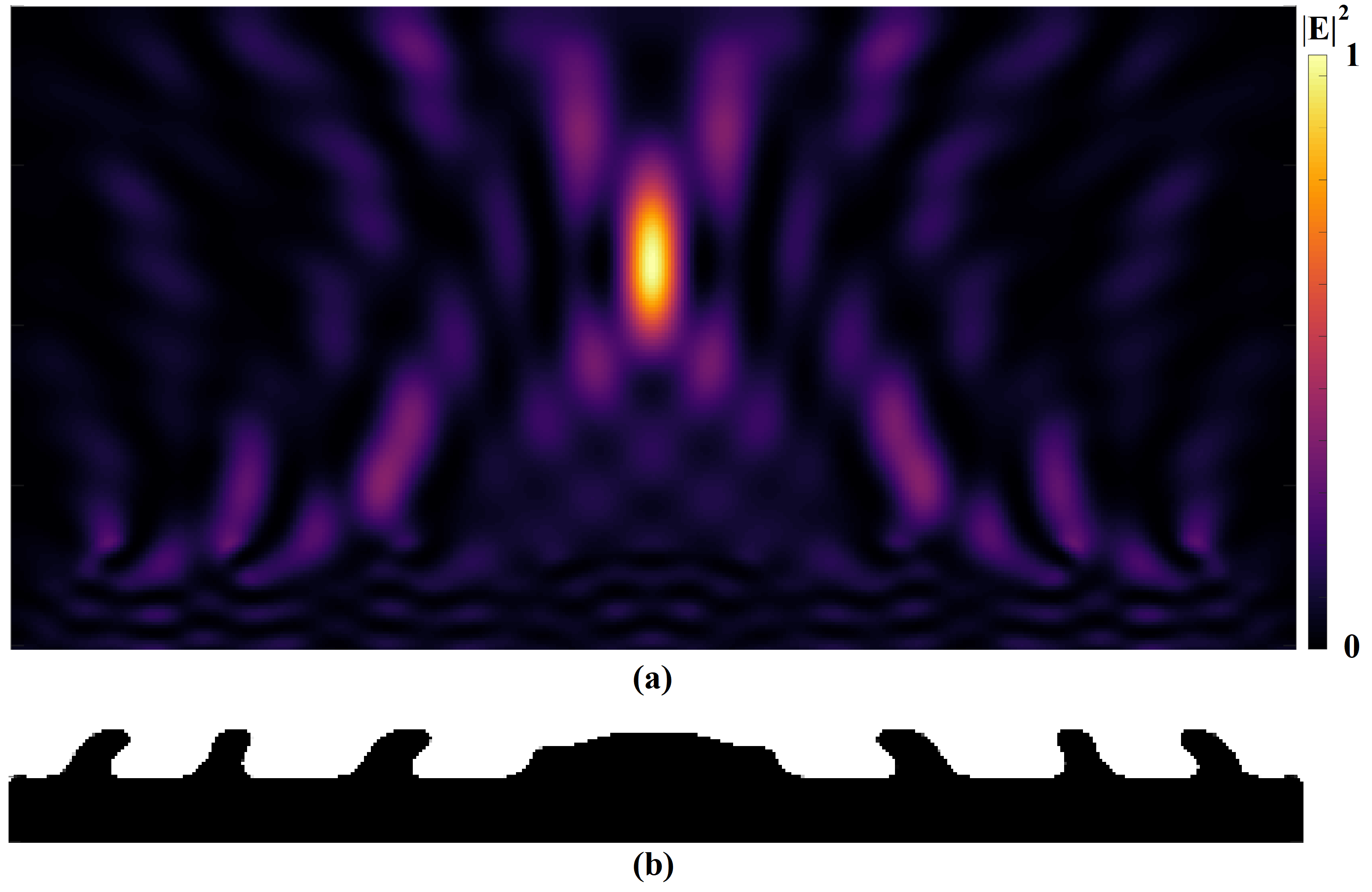} \caption{\textbf{(a)} Max-normalized $\vert \textbf{E} \vert^2$-field in $\Omega$. \textbf{(b)} Metalens design, $\varepsilon_r=3.0$ (black) and $\varepsilon_r=1.0$ (white).} \label{FIG:SOLUTION_1}
	}
\end{figure}

\noindent The final binarized design is shown in Fig.~\ref{FIG:SOLUTION_1}b with black(white) representing solid(air). Figure~\ref{FIG:SOLUTION_1}a shows the $\vert E_z \vert^2$-field resulting from exciting the metalens in Fig.~\ref{FIG:SOLUTION_1}b for the targeted incident field, demonstrating the focussing effect at the targeted focal spot. The numerical aperture of the metalens is NA $\approx 0.92$ and the transmission efficiency is $T_A \approx 0.87$ computed as the power propagating through the lens relative to the power incident on the lens.

\subsection{\textcolor{black}{Solving the same design problem with different resolutions}}

\textcolor{black}{For various reasons, such as performing a mesh-convergence study, it may be necessary to solve the same physical model problem using different mesh resolutions. This may be done with} \verb|top200EM| \textcolor{black}{by multiplying the following inputs by an integer scaling factor: The number of finite elements in each spatial direction,} \verb|nElX| \textcolor{black}{and} \verb|nElY|, \textcolor{black}{the wavelength} \verb|lambda| \textcolor{black}{and the filter radius} \verb|fR| \textcolor{black}{and dividing the hard-coded value of} \verb|phys.scale| \textcolor{black}{in the code by the same factor.}

\subsection{The Effect of Filtering}

\begin{figure}[h!]
	\centering
	{
		\includegraphics[width=0.95\textwidth]{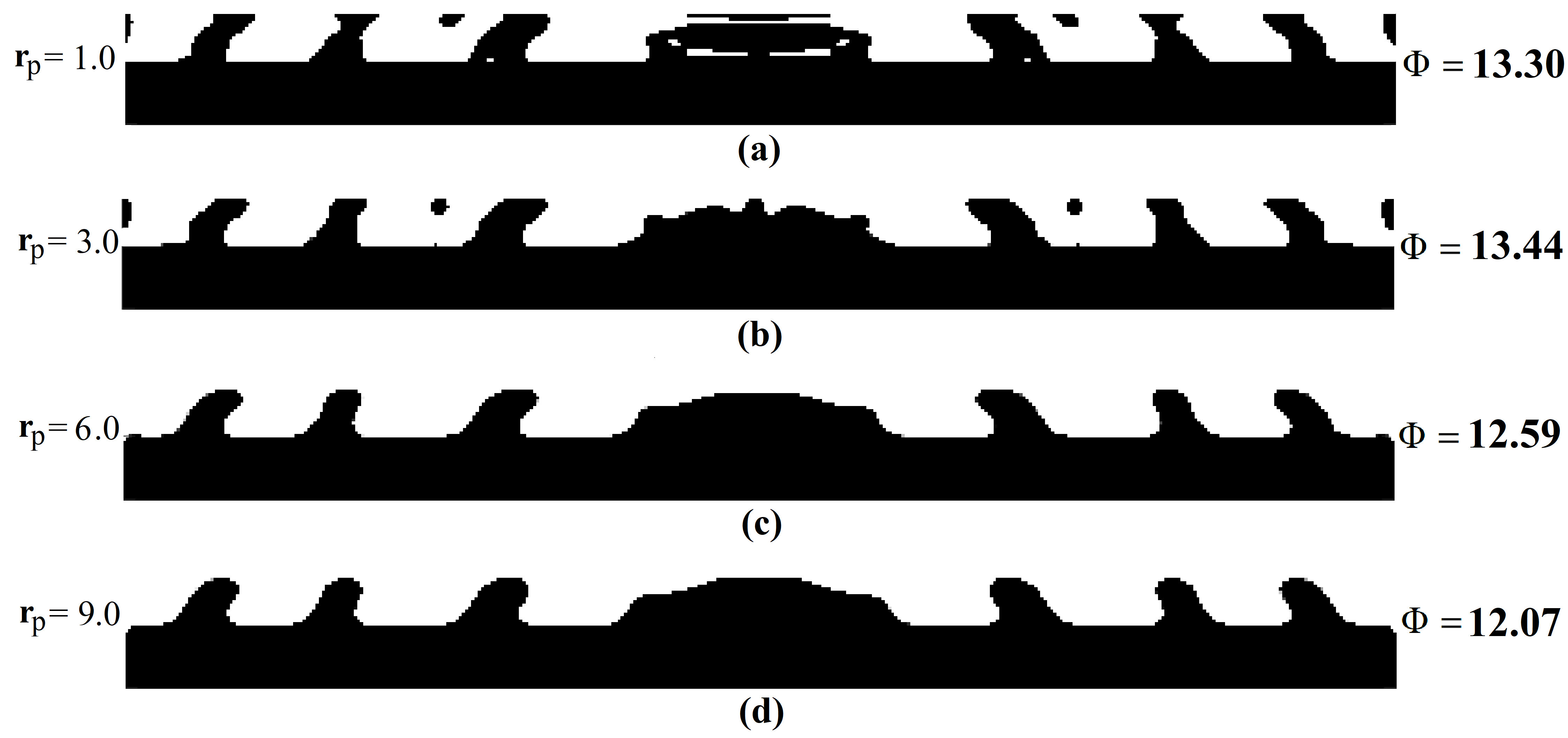} \caption{Metalens designs obtained \textbf{(a)}  without filtering $(r_f = 1)$, \textbf{(b)} using a filter radius of $r_f = 3.0$, \textbf{(c)} using a filter radius of $r_f = 6.0$, \textbf{(d)} using a filter radius of $r_f = 9.0$. These results illustrate the effect of applying the cone-shaped filter to the design field as part of the optimization process. \label{FIG:SOLUTION_2}}
	}
\end{figure}

Next, we demonstrate the effect of applying the filtering step \cite{LAZAROV_ET_AL_2016} by changing the filter radius and designing four metalenses (See Fig.~\ref{FIG:SOLUTION_2}) using \verb|top200EM|. Again, the model-domain size and indices for the design domain are defined first,

\begin{itemize}
	\item[] 
\begin{verbatim}
% DESIGN FIELD INDICES
DomainElementsX = 400;
DomainElementsY = 200;
DesignThicknessElements = 15;
DDIdx = repmat([1:DomainElementsY:DomainElementsX*DomainElementsY],...
DesignThicknessElements,1);
DDIdx = DDIdx+repmat([165:165+DesignThicknessElements-1]',1,DomainElementsX);
\end{verbatim}
\end{itemize}

\noindent Then, the optimization \textcolor{black}{problem is solved} with the four filtering radii \verb|fR|$=r_f \in \lbrace 1.0, 3.0, 6.0, 9.0 \rbrace$,

\begin{itemize}
	\item[] 
\begin{verbatim}
[DVs,obj]=top200EM([200,80],DDIdx,DomainElementsX,DomainElementsY,...
0.5,3.0,35,fR,200);
\end{verbatim}
\end{itemize}

\noindent Looking at the four final binarized designs in Fig.~\ref{FIG:SOLUTION_2}, it is clearly observed that as $r_f$ is increased the features in the designs grow. When no filtering is applied (Fig.~\ref{FIG:SOLUTION_2}a) single-pixel sized features are observed. Such features may be problematic from a numerical-modelling point of view as well as being detrimental to fabrication. 

\textcolor{black}{It is noted that without filtering TopOpt is more prone to identify a local minimum, which performs worse than ones identified with filtering. In other words, filtering tends to have a convexifying effect, as long as the filter size is not too big.}

\section{Modifying the Code} \label{SEC:ADVANCED}

There exists a vast amount of auxiliary tools developed to extend the applicability of density-based Topology Optimization across a wide range of different problems and physics. The following sections provide examples of how simple some of these tools are to implement in \verb|top200EM|.

\subsection{Plasmonics} \label{SEC:PLASMONIC_INTERPOLATION}

In recent work \cite{CHRISTIANSEN_VESTER_2019} it was demonstrated that \textcolor{black}{using a refractive index and extinction cross section based} non-linear material interpolation yielded significantly improved results when designing Au, Ag and Cu nano-particles for localized field enhancement, with recent applications to enhanced upconversion \cite{CHRISTIANSEN_SEMSC_2020}. thermal emission \cite{KUDYSHEV_ET_AL_2020} and Raman scattering \cite{CHRISTIANSEN_OE_2020}. \textcolor{black}{This interpolation function avoids artificial resonances in connection with the transition from positive to negative $\varepsilon$ values and reads,}

\begin{eqnarray*}
\varepsilon(x) = (n(x)^2 - \kappa(x)^2) - \text{i} (2 n(x) \kappa(x)), \ n(x) = n_{\text{M}_1} + x \ (n_{\text{M}_2} - n_{\text{M}_1}), \ \kappa(x) = \kappa_{\text{M}_1} + x \ (\kappa_{\text{M}_2} - \kappa_{\text{M}_1}).
\end{eqnarray*}

\noindent Here $n$ and $\kappa$ denote the refractive index and extinction cross section, respectively. The subscripts $M_1$ and $M_2$ denote the two materials being interpolated. \\

\noindent The interpolation scheme is straight forward to implement in the code as follows. \\

\noindent First the following lines of code, 

\begin{itemize}
	\item[] 
\begin{verbatim}
function [A,dAdx] = MATERIAL_INTERPOLATION(eps_r,x,1.0)
A = 1 + x*(eps_r-1) - 1i * alpha_i * x .* (1 - x); % Interpolation
dAdx = (eps_r-1)*(1+0*x) - 1i * alpha_i * (1 - 2*x); % Derivative of interpolation
end
\end{verbatim}
\end{itemize}

\noindent are replaced with, 

\begin{itemize}
	\item[] 
\begin{verbatim}
function [A,dAdx] = MATERIAL_INTERPOLATION(n_r,k_r,x) 
n_eff = 1 + x*(n_r-1);
k_eff = 0 + x*(k_r-0);
A = (n_eff.^2 - k_eff.^2) - 1i*(2.*n_eff.*k_eff);
dAdx = 2*n_eff*(n_r-1)-2*k_eff*(k_r-1)-1i*(2*(n_r-1)*k_eff+(2*n_eff*(k_r-1)));
end
\end{verbatim}
\end{itemize}

\noindent where for simplicity it is assumed that $M_1$ is air, i.e. $n_{\text{M}_1} = 1.0$ and $\kappa_{\text{M}_1} = 0.0$. \\

\noindent Second, the scalar input parameter \verb|eps_r| is changed to a two-valued 1D-array \verb|nk_r|. \\
 
\noindent Third, the line, 

\begin{itemize}
	\item[] 
\begin{verbatim}
phy.eps_r = eps_r; % Relative permittivity
\end{verbatim}
\end{itemize}

\noindent is replaced with, 

\begin{itemize}
	\item[] 
\begin{verbatim}
phy.nk_r = nk_r; % Refractive index and extinction coefficient 
\end{verbatim}
\end{itemize}

\noindent and fourth, the call to the material interpolation function is changed from,

\begin{itemize}
	\item[] 
	\begin{verbatim}
[A,dAdx] = MATERIAL_INTERPOLATION(phy.eps_r,dFPST,1.0); 
	\end{verbatim}
\end{itemize}

\noindent to, 

\begin{itemize}
	\item[] 
	\begin{verbatim}
[A,dAdx] = MATERIAL_INTERPOLATION(phy.nk_r(1),phy.nk_r(2),dFPST); 
	\end{verbatim}
\end{itemize}

\subsection{Excitation} \label{SEC:INCIDENT_FIELD}

The excitation considered in the \textcolor{black}{baseline} problem may be changed straightforwardly. As an example, the boundary at which the incident field enters the domain may be changed from the bottom to the top boundary as follows. \\

\noindent First, the index set controlling where the boundary condition is imposed in the vector, $\textbf{F}$(=\verb|F|), is changed by moving \verb|dis.iRHS = TMP;| from line 157 to above line 151. \\

\noindent Second, the values stored in \verb|F| are changed to account for the propagation direction of the wave by replacing,

\begin{itemize}
	\item[] 
\begin{verbatim}
F(dis.iRHS(1,:)) = F(dis.iRHS(1,:))+1i*waveVector;
F(dis.iRHS(2,:)) = F(dis.iRHS(2,:))+1i*waveVector; 
\end{verbatim}
\end{itemize}

\noindent with, 

\begin{itemize}
	\item[] 
\begin{verbatim}
F(dis.iRHS(1,:)) = F(dis.iRHS(1,:))-1i*waveVector; 
F(dis.iRHS(2,:)) = F(dis.iRHS(2,:))-1i*waveVector;
\end{verbatim}
\end{itemize}

\subsection{Designing a Metallic Reflector}

By introducing the changes presented in Sec.~\ref{SEC:PLASMONIC_INTERPOLATION} and Sec.~\ref{SEC:INCIDENT_FIELD}, one may design a metallic reflector using \verb|topEM200|. \\

\begin{figure}[h!]
	\centering
	{
		\includegraphics[width=0.85\textwidth]{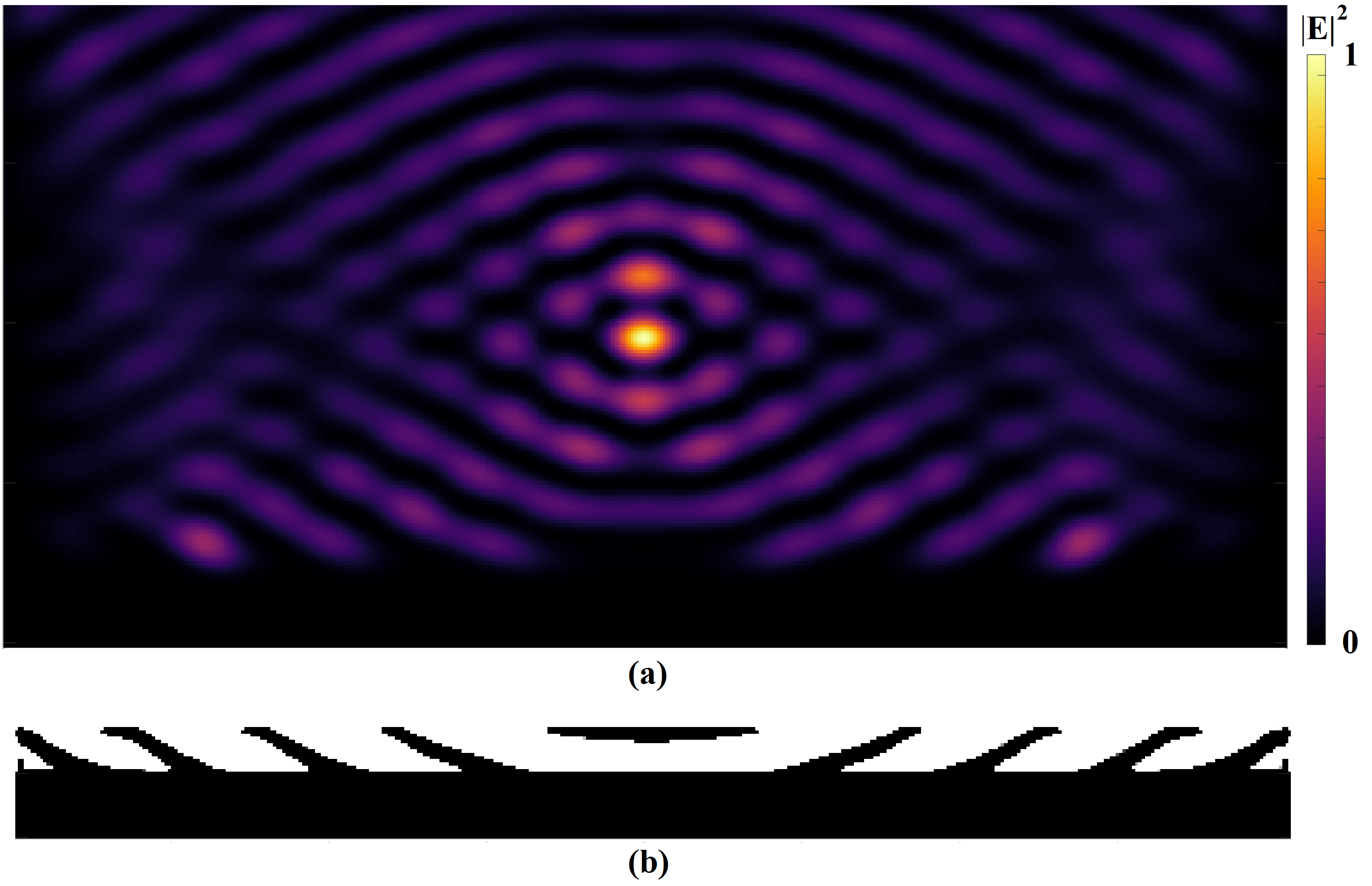} \caption{\textbf{(a)} Max-normalized $\vert \textbf{E} \vert^2$-field in $\Omega$. \textbf{(b)} Metallic reflector design (black) in air background (white).} \label{FIG:METAL_REFLECTOR}
	}
\end{figure}

\noindent First, we set the domain dimensions and design thickness as in the previous examples,

\begin{itemize}
	\item[] 
	\begin{verbatim}
% DESIGN FIELD INDICES
DomainElementsX = 400;
DomainElementsY = 200;
DesignThicknessElements = 15;
DDIdx = repmat([1:DomainElementsY:DomainElementsX*DomainElementsY],...
DesignThicknessElements,1);
DDIdx = DDIdx+repmat([165:165+DesignThicknessElements-1]',1,DomainElementsX);
	\end{verbatim}
\end{itemize}

\noindent Second, we solve the optimization problem by executing the command,

\begin{itemize}
	\item[] 
	\begin{verbatim}
[DVs,obj]=top200EM([200,100],DDIdx,DomainElementsX,DomainElementsY,...
0.5,[1.9,1.5],35,3.0,200);
	\end{verbatim}
\end{itemize}

\noindent \textbf{Note:} For simplicity we selected the following values for the refractive index, $n=1.9$(=\verb|nk_r(1)|), and extinction cross section, $\kappa=1.5$(=\verb|nk_r(2)|).\footnote{These values corresponds to values for gold at $\lambda = 350$ nm. I.e. one may think of this choice of material parameters as an unspoken rescaling of space by a factor of 10 i.e. changing the element size to pixels of 10 nm by 10 nm rather than of 1 nm by 1 nm.} \\

\noindent Considering the final binarized reflector design in Fig.~\ref{FIG:METAL_REFLECTOR}b, one can interpret the design as the well-known parabolic reflector broken into pieces to fit the spatially-limited design domain. From the max-normalized $\vert \textbf{E} \vert^2$-field presented in Fig.~\ref{FIG:METAL_REFLECTOR}a the focussing effect of the reflector is clearly observed. 

\subsection{Linking Design Variables} \label{SEC:LINKING_VARIABLES}

Certain fabrication techniques limit the allowable geometric variations in a design. For example, optical-projection lithography and electron-beam lithography restricts variations in design geometries to two-dimensional patterns, which can then be extruded in the out-of-plane direction. It is straight forward to introduce such a geometric restriction using TopOpt by linking design variables and sensitivities across elements. In \verb|top200EM| the design field may be restricted to only exhibit in-plane \textcolor{black}{(x-direction)} variations as follows. \\

\noindent First, the code-line representing the values of the design variables,

\begin{itemize}
	\item[] 
\begin{verbatim}
dVs(length(dis.dVElmIdx(:))) = dVini; % Design variables
\end{verbatim}
\end{itemize}

\noindent is replaced with,

\begin{itemize}
	\item[] 
\begin{verbatim}
dVs(1:nElX) = dVini; % Design variables for 1D design 
\end{verbatim}
\end{itemize}

\noindent Second, the code-line transferring the design variables to the elements in the physics model, 

\begin{itemize}
	\item[] 
\begin{verbatim}
dFP(dis.dVElmIdx(:)) = dVs; % Design variables inserted in design field
\end{verbatim}
\end{itemize}

\noindent is replaced with,

\begin{itemize}
	\item[] 
\begin{verbatim}
nRows=length(dis.dVElmIdx(:))/dis.nElX; % Number of rows in the 1D design
dFP(dis.dVElmIdx) = repmat(dVs,1,nRows)'; % Design variables inserted in design field
\end{verbatim}
\end{itemize}

\noindent and finally the code-line representing individual element sensitivities,

\begin{itemize}
	\item[] 
\begin{verbatim}
sensFOM = sensFOM(dis.dVElmIdx);
\end{verbatim}
\end{itemize}

\noindent is replaced with,

\begin{itemize}
	\item[] 
\begin{verbatim}
sensFOM = sensFOM(dis.dVElmIdx);
sensFOM = sum(sensFOM,1); % Correcting sensitivities for 1D design
\end{verbatim}
\end{itemize}

\noindent which sums the sensitivity contributions from the linked elements.

\subsection{Designing a 1D metalens}

By introducing the changes listed in Sec.~\ref{SEC:LINKING_VARIABLES} \verb|topEM200| can be used to design metasurfaces of fixed height with in-plane variations as follows. Again we define the domain dimensions as,

\begin{itemize}
	\item[] 
	\begin{verbatim}
% DESIGN FIELD INDICES
DomainElementsX = 400;
DomainElementsY = 200;
DesignThicknessElements = 15;
DDIdx = repmat([1:DomainElementsY:DomainElementsX*DomainElementsY],...
DesignThicknessElements,1);
DDIdx = DDIdx+repmat([165:165+DesignThicknessElements-1]',1,DomainElementsX);
	\end{verbatim}
\end{itemize}

\noindent Followed by the execution of the command,

\begin{itemize}
	\item[] 
	\begin{verbatim}
[DVs,obj]=top200EM([200,80],DDIdx,DomainElementsX,DomainElementsY,...
0.5,3.0,35,3.0,200);
	\end{verbatim}
\end{itemize}

\begin{figure}[h!]
	\centering
	{
		\includegraphics[width=0.85\textwidth]{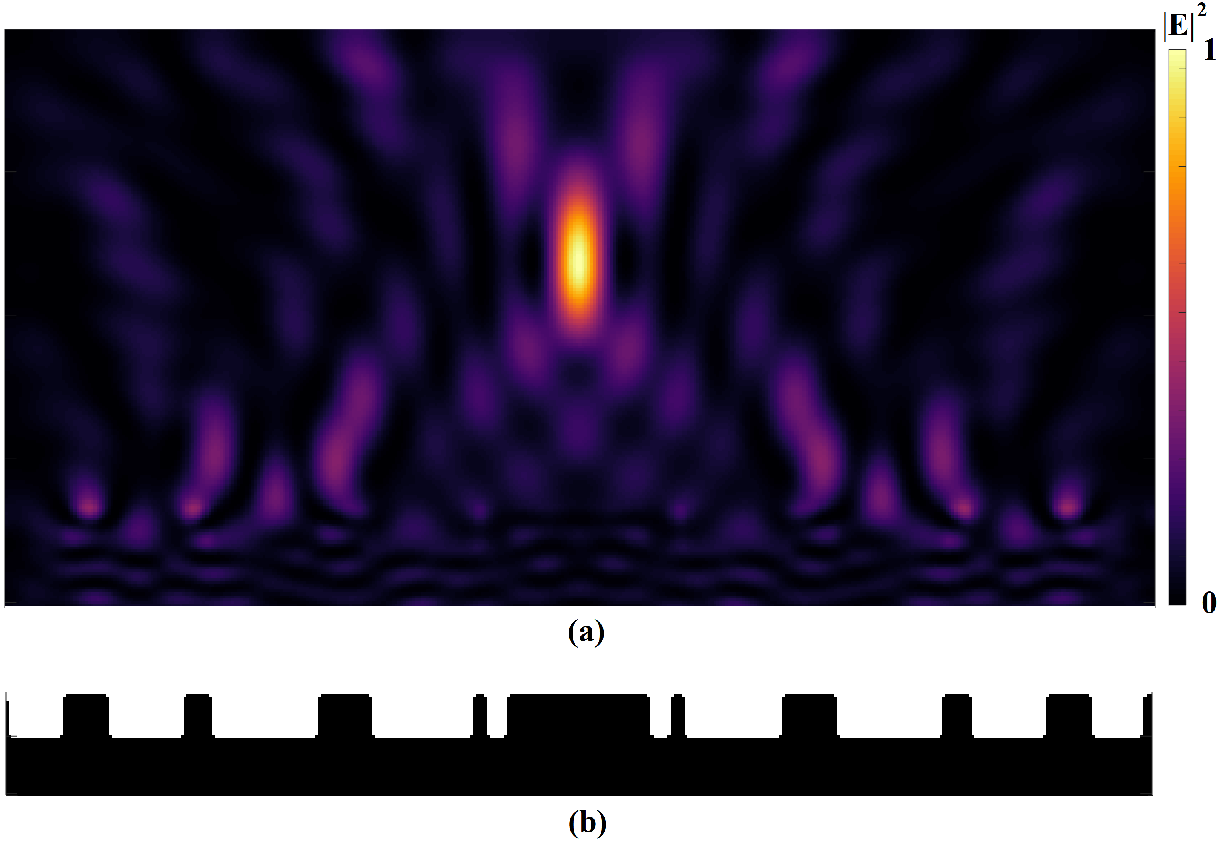} \caption{\textbf{(a)} Max-normalized $\vert \textbf{E} \vert^2$-field in $\Omega$. \textbf{(b)} Metalens design restricted to one dimensional variations, $\varepsilon_r=3.0$ (black) and $\varepsilon_r=1.0$ (white).} \label{FIG:1D_DESIGN}
	}
\end{figure}

\noindent The final binarized design, resulting from solving the design problem, is shown in Fig.~\ref{FIG:1D_DESIGN}b. Here it is clear to see that the design is now restricted to vary only in the x-direction. It is worth noting that the design is still filtered in both spatial directions, hence the corners of the design appear rounded in Fig.~\ref{FIG:1D_DESIGN}b. The max-normalized $\vert \textbf{E} \vert^2$-field presented in Fig.~\ref{FIG:METAL_REFLECTOR}a demonstrates the focussing effect of the lens at the targeted point in the modelling domain, which, due to the reduced design freedom, is not as high as in the un-restricted case \textcolor{black}{(Sec. \ref{SEC:DESIGN_METALENS})}.

\subsection{Continuation of Threshold Sharpness}

For some design problems in photonics and plasmonics, e.g. \cite{CHRISTIANSEN_VESTER_2019,CHRISTIANSEN_OE_2020,WANG_ET_AL_2011}, intermediate values may be present in the final optimized design, i.e. $\bar{\tilde{\boldmath{\xi}}}_{\mathrm{final}} \in ]0,1[$, despite the applied penalization scheme, as they prove beneficial to optimizing the FOM. However, (in most cases) intermediate values hold no physical meaning and it is therefore not possible to realize designs containing such intermediate values experimentally. \textcolor{black}{A way} to promote that (almost) no design variables take intermediate values in the final design, is by using a continuation scheme for the threshold sharpness, gradually increasing it until an (almost) pure 0/1-design is achieved, ensuring that the optimized designs are physically realizable. \\

\noindent To implement the continuation scheme in the code, replace the following lines,

\begin{itemize}
	\item[] 
	\begin{verbatim}
	% SOLVE DESIGN PROBLEM USING MATLAB BUILT-IN OPTIMIZER: FMINCON
	FOM = @(dVs)OBJECTIVE_GRAD(dVs,dis,phy,filThr);
	[dVs,~] = fmincon(FOM,dVs(:),[],[],[],[],LBdVs,UBdVs,[],options);
	\end{verbatim}
\end{itemize}

\noindent with,

\begin{itemize}
	\item[] 
	\begin{verbatim}
	while filThr.beta<betaMax % Thresholding sharpness bound
		% SOLVE DESIGN PROBLEM USING MATLAB BUILT-IN OPTIMIZER: FMINCON
		FOM = @(dVs)OBJECTIVE_GRAD(dVs,dis,phy,filThr);
		[dVs,~] = fmincon(FOM,dVs(:),[],[],[],[],LBdVs,UBdVs,[],options);
		filThr.beta = betaInc * filThr.beta; % Increasing thresholding sharpness
	end
	\end{verbatim}
\end{itemize}

\noindent and select suitable values for \verb|betaMax| and \verb|betaInc|. These values are problem dependent and some experimentation may be required to identify the best values for a given problem. For the metalens design example in Sec.~\ref{SEC:DESIGN_METALENS}, when considering high material contrast, i.e. large values of \verb|eps_r|, the values \verb|betaMax = 20.0| and \verb|betaInc = 1.5| have been found to work well.

\section{What about genetic algorithms?} \label{SEC:GA_VS_TOPOPT}

For a wide range of inverse design problems, where gradient-based optimization methods are applicable\footnote{Inverse design problems where it is possible to compute the sensitivities of the FOM (and any constraints), e.g. using adjoint sensitivity analysis, a pre-requisite for using gradient-based methods.}, they have been found to severely outperform non-gradient-based methods, such as genetic algorithms (GA) \cite{GOLDBERG_1989}. This, both in terms of the computational effort required to identify a local optimum for the FOM and, by direct extension, in terms of the number of design degrees of freedom it is feasible to consider (a difference of many orders of magnitude \cite{AAGE_ET_AL2017}). Further, in many cases, gradient-based methods are able to identify better local optima for the FOM \cite{SIGMUND_2011}. 

In the following, we provide a demonstration of these claims, by comparing the solution of a metalens design problem obtained using the gradient-based \verb|top200EM| to the solution obtained using MATLABs built in genetic algorithm \verb|ga|. \\

\noindent Readers may perform their own comparisons by rewriting \verb|top200EM| to utilize \verb|ga| instead of \verb|fmincon| as follows. \\

\noindent First, a (near perfect) 0/1-design is ensured by changing the thresholding strength from,

\begin{itemize}
	\item[] 
	\begin{verbatim}
	filThr.beta = 5; % Thresholding sharpness
	\end{verbatim}
\end{itemize}

\noindent to, 

\begin{itemize}
	\item[] 
	\begin{verbatim}
	filThr.beta = 1e5; % Thresholding sharpness
	\end{verbatim}
\end{itemize}

\noindent Second, \verb|ga| is used instead of \verb|fmincon| by replacing the following lines of codes,

\begin{itemize}
	\item[] 
	\begin{verbatim}
	options = optimoptions('fmincon','Algorithm','interior-point',...
	'SpecifyObjectiveGradient',true,'HessianApproximation','lbfgs',...
	'Display','off','MaxIterations',maxItr,'MaxFunctionEvaluations',maxItr);
	
	% SOLVE DESIGN PROBLEM USING MATLAB BUILT-IN OPTIMIZER: FMINCON
	FOM = @(dVs)OBJECTIVE_GRAD(dVs,dis,phy,filThr);
	[dVs,~] = fmincon(FOM,dVs(:),[],[],[],[],LBdVs,UBdVs,[],options);
	\end{verbatim}
\end{itemize}

\noindent with, 

\begin{itemize}
	\item[] 
	\begin{verbatim}
	options = optimoptions('ga','MaxGenerations',maxItr,'Display','iter'); 
	
	% SOLVE DESIGN PROBLEM USING MATLAB BUILT-IN OPTIMIZER: GA
	FOM = @(dVs)OBJECTIVE(dVs,dis,phy,filThr);
	rng(1,'twister'); % SETTING RANDOM SEED TO ENSURE REPRODUCTION OF RESULT
	[dVs,~,~,~] = ga(FOM,length(dVs),[],[],[],[],LBdVs,UBdVs,[],[],options);
	\end{verbatim}
\end{itemize}

\noindent where the command \verb|rng(1,'twister')| fixes the random seed for reproducibility.

\noindent Third, the computation of the gradient of the FOM is removed by changing,

\begin{itemize}
	\item[] 
	\begin{verbatim}
	function [FOM,sensFOM] = OBJECTIVE_GRAD(dVs,dis,phy,filThr)
	\end{verbatim}
\end{itemize}

\noindent to,

\begin{itemize}
	\item[] 
	\begin{verbatim}
	function [FOM] = OBJECTIVE(dVs,dis,phy,filThr)
	\end{verbatim}
\end{itemize}

\noindent Fourth, deleting the following lines in \verb|OBJECTIVE|,

\begin{itemize}
	\item[] 
	\begin{verbatim}
	% ADJOINT RIGHT HAND SIDE 
	AdjRHS = P*(2*real(Ez) - 1i*2*imag(Ez));
	
	% SOLVING THE ADJOING SYSTEM: S.' * AdjLambda = AdjRHS
	AdjLambda = (Q1.') * ((L.')\((U.')\((Q2.') * (-1/2*AdjRHS)))); % Solving
	
	% COMPUTING SENSITIVITIES
	dis.vDS = reshape(-phy.k^2*dis.MEM(:)*(dAdx(:).'),16*dis.nElX*dis.nElY,1); 
	DSdx = sparse(dis.iElFull,dis.jElFull,dis.vDS); % Constructing dS/dx
	DSdxMulV = DSdx * Ez(dis.idxDSdx); % Computing dS/dx * Field values
	DsdxMulV = sparse(dis.iElSens,dis.jElSens,DSdxMulV);
	sens = 2*real(AdjLambda(dis.idxDSdx).' * DsdxMulV); % Computing sensitivites
	sens = full(reshape(sens,dis.nElY,dis.nElX));
	
	% FILTERING SENSITIVITIES
	DdFSTDFS = DERIVATIVE_OF_THRESHOLD( dFPS, filThr.beta, filThr.eta);
	sensFOM = DENSITY_FILTER(filThr.filKer,filThr.filSca,sens,DdFSTDFS);
	
	% EXTRACTING SENSITIVITIES FOR DESIGNABLE REGION
	sensFOM = sensFOM(dis.dVElmIdx);
	
	% FMINCON DOES MINIMIZATION
	sensFOM = -sensFOM(:);
	
	\end{verbatim}
\end{itemize}

\noindent and finally changing the code evaluating the final design,

\begin{itemize}
	\item[] 
	\begin{verbatim}
	% FINAL BINARIZED DESIGN EVALUATION
	filThr.beta = 1000; 
	disp('Black/white design evaluation:');
	[FOM,~] = OBJECTIVE(dVs(:),dis,phy,filThr); 
	\end{verbatim}
\end{itemize}

\noindent to, 

\begin{itemize}
	\item[] 
	\begin{verbatim}
	% FINAL BINARIZED DESIGN EVALUATION
	filThr.beta = 1e5; 
	disp('Black/white design evaluation:');
	[FOM,~] = OBJECTIVE(dVs(:),dis,phy,filThr); 
	\end{verbatim}
\end{itemize}

\noindent \textbf{Note:} The code-lines plotting the design in \verb|OBJECTIVE| can be removed to reduce wallclock time. \\

\noindent In order to reduce the computational effort required to reproduce the following example, we consider a problem in a smaller spatial domain, considering a shorter wavelength and fewer degrees of freedom than in the previous examples. This is done by defining a model problem considering 1000 design degrees of freedom as,

\begin{itemize}
	\item[] 
	\begin{verbatim}
	% DESIGN FIELD INDICES
	DomainElementsX = 100;
	DomainElementsY = 50;
	DesignThicknessElements = 10;
	DDIdx = repmat([1:DomainElementsY:DomainElementsX*DomainElementsY],...
	DesignThicknessElements,1);
	DDIdx = DDIdx+repmat([35:35+DesignThicknessElements-1]',1,DomainElementsX);
	\end{verbatim}
\end{itemize}

Followed by the solution of the design problem using the modified GA-based code,

\begin{itemize}
	\item[] 
	\begin{verbatim}
	[dVs_GA,FOM_GA]=top200EMGA([50,10],DDIdx,DomainElementsX,DomainElementsY,...
	0.5,3.0,20,3.0,500);
	\end{verbatim}
\end{itemize}

and then using the original gradient-based code, 

\begin{itemize}
	\item[] 
	\begin{verbatim}
	[dVs,FOM]=top200EM([50,10],DDIdx,DomainElementsX,DomainElementsY,...
	0.5,3.0,20,3.0,500);
	\end{verbatim}
\end{itemize}

\begin{figure}[h!]
	\centering
	{
		\includegraphics[width=1.0\textwidth]{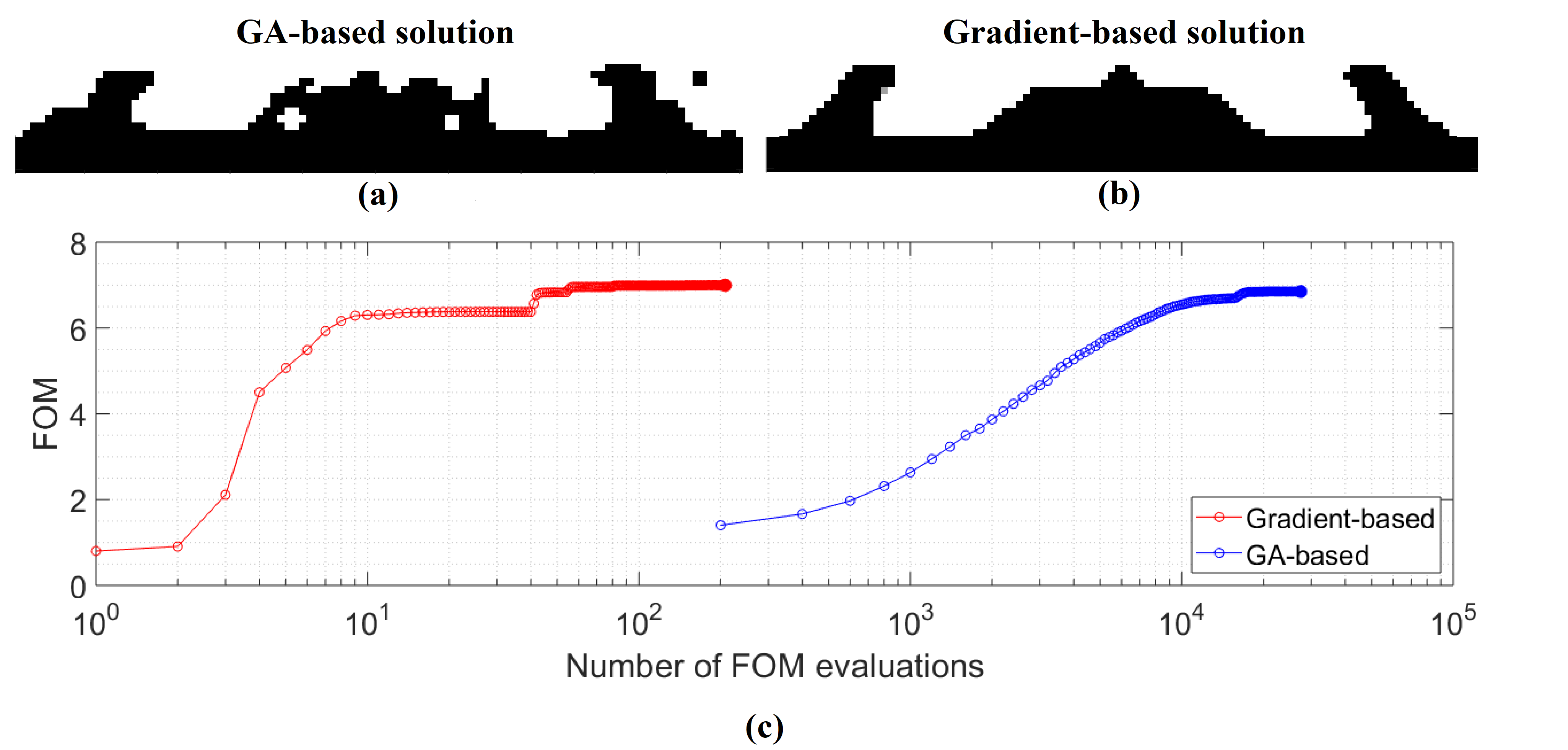} \caption{\textbf{(a)} Design obtained using the GA-based method. \textbf{(b)} Design obtained using the gradient-based method. $\varepsilon_r=3.0$ (black) and $\varepsilon_r=1.0$ (white). \textbf{(c)} Convergence graph. Remark that each design iteration for the GA-based code requires 200 solutions of the physics model equation, while it requires 1 solution of the physics model equation and 1 solution of the adjoint equation for the gradient-based code! \label{FIG:GA_VERSUS_GRADIENT}}
	}
\end{figure}

The binarized designs obtained using \verb|top200EMGA| and \verb|top200EM| are presented in Fig.~\ref{FIG:GA_VERSUS_GRADIENT}a and Fig.~\ref{FIG:GA_VERSUS_GRADIENT}b, respectively. The evolution of the FOM value as function of the number of FOM evaluations is plotted in Fig.~\ref{FIG:GA_VERSUS_GRADIENT}c for both methods.

The GA-based method uses 137 design iterations (27,600 FOM evaluations) to identify a solution. In contrast the gradient-based method uses 208 iterations (208 FOM evaluation) to identify a superior solution. Evaluating the final binarized designs the GA-based solution has a FOM value of $\Phi \approx 6.85$ while the gradient-based solution has an $\approx 18\%$ better FOM value of $\Phi \approx 8.07$.

Crucially, the GA-based method uses 200 FOM evaluations per design iteration to drive the optimization process. This results in a total of 27.600 FOM evaluations for solving the design problem using the GA-based method, which corresponds to 27.600 solutions of the physics model equation, i.e. eq.~(\ref{EQN:FORWARD_PROBLEM}). In contrast, when solving the design problem using the gradient-based method a total of 208 FOM evaulations are performed, corresponding to 208 solution of the physics model equation plus 208 solutions of the adjoint system (eq.~(\ref{EQN:ADJOINT_LINEAR_SYSTEM})), which for the baseline example is almost free due to the reuse of the LU-factorization. 

Hence, in our example the identification of a local optimum for the FOM using the gradient-based method required $\approx 1.0\%$ of the computational effort spend by the GA-based method. A ratio, which only becomes more pronounced as the size of the design space grows. 

This example is based on standard settings for MATLAB’s \verb|ga| optimizer and can surely be improved and refined. However, this will not change the basic conclusion that non-gradient-based approaches are unsuitable for large-scale TopOpt problems. A more thorough discussion of non-gradient-based methods in TopOpt can be found in Ref. \cite{SIGMUND_2011}. Readers are invited to form their own conclusions by extending the example and codes provided here.

\section{Conclusion}

We \textcolor{black}{have} presented a simple finite element based MATLAB code (downloadable from \url{https://www.topopt.mek.dtu.dk}) for TopOpt-based inverse design of photonic structures. We \textcolor{black}{have} provided examples of how to use the code, as well as a set of suggestions for code extensions enabling it to handle metallic structures, different model excitations, linked design variables and introducing a continuation scheme for the threshold sharpness designed to promote 0/1-designs. \textcolor{black}{Finally, we have demonstrated the superiority of a gradient-based method over a genetic-algorithm-based method when performing inverse design in photonics}. 

The code can be used for educational purposes as is, and is otherwise meant to serve as a starting point for the reader to develop software to handle their more advanced research applications within photonics. For simplicity and computational speed, the code treats problems in two spatial dimensions, however it is directly extendable to three spatial dimensions by modifying the finite element matrices, boundary conditions and index sets appropriately. 

\appendix

\section{Adjoint Sensitivity Analysis} \label{APN:ADJOINT_SENSITIVITY_ANALYSIS}

The expression for $\frac{\mathrm{d} \boldmath{\Phi}}{\mathrm{d} \bar{\tilde{\boldmath{\xi}}}_k}$ in eq.~(\ref{EQN:SENSITIVITES}) and the expression in eq.~(\ref{EQN:ADJOINT_LINEAR_SYSTEM}) may be derived as follows. First, zero is added to $\boldmath\Phi$ twice,

\begin{eqnarray} \label{EQN:ADJOINT_DERIVATIVE_DERIVATION_STEP_1}
\tilde{ \boldmath{\Phi}} =  \boldmath{\Phi} + \boldmath{\lambda}^{\mathrm{T}} \left( \textbf{S} \textbf{E}_z - \textbf{F} \right)  + \boldmath{\lambda}^\dagger \left( \textbf{S}^{*} \textbf{E}_z^{*} - \textbf{F}^{*} \right)
\end{eqnarray}

\noindent where $\left( \textbf{S} \textbf{E}_z - \textbf{F} \right) = 0$ and $\boldmath{\lambda}$ is a vector of nodal complex Lagrange multipliers, also called the adjoint variables. Second, one takes the derivative of $\tilde{ \boldmath{\Phi}}$ with respect to $\bar{\tilde{\boldmath{\xi}}}_k$ and exploits that, for the optimization problem in eq.~(\ref{EQN:OPTIMIZATION_PROBLEM}), $\boldmath\Phi$ does not depend explicitly on $\bar{\tilde{\boldmath{\xi}}}_k$ and neither $\boldmath{\lambda}$ nor $\textbf{F}$ depend on $\bar{\tilde{\boldmath{\xi}}}_k$ at all, yielding,

\begin{eqnarray} \label{EQN:ADJOINT_DERIVATIVE_DERIVATION_STEP_2}
\frac{\mathrm{d} \tilde{\boldmath{\Phi}}}{\mathrm{d} \bar{\tilde{\boldmath{\xi}}}_k} =  \frac{\partial \boldmath{\Phi} }{ \partial \textbf{E}_{z,\Re} } \frac{\partial \textbf{E}_{z,\Re} }{\partial \bar{\tilde{\boldmath{\xi}}}_k } + \frac{\partial \boldmath{\Phi} }{ \partial \textbf{E}_{z,\Im} } \frac{\partial \textbf{E}_{z,\Im} }{\partial \bar{\tilde{\boldmath{\xi}}}_k } + \boldmath{\lambda}^{\mathrm{T}} \left( \frac{\partial \textbf{S}}{\mathrm{d} \bar{\tilde{\boldmath{\xi}}}_k} \textbf{E}_z + \textbf{S}\left( \frac{\partial \textbf{E}_{z,\Re} }{\partial \bar{\tilde{\boldmath{\xi}}}_k } + \mathrm{i}  \frac{\partial \textbf{E}_{z,\Im} }{\partial \bar{\tilde{\boldmath{\xi}}}_k } \right)  \right) \\ 
+ \boldmath{\lambda}^\dagger \left( \frac{\partial \textbf{S}^{*}}{\partial \bar{\tilde{\boldmath{\xi}}}_k} \textbf{E}_z^{*} + \textbf{S}^{*}\left( \frac{\partial \textbf{E}_{z,\Re} }{\partial \bar{\tilde{\boldmath{\xi}}}_k } - \mathrm{i}  \frac{\partial \textbf{E}_{z,\Im} }{\partial \bar{\tilde{\boldmath{\xi}}}_k } \right) \right), \nonumber
\end{eqnarray}

\noindent Collecting terms including $\frac{\partial \textbf{E}_{z,\Re} }{\partial \bar{\tilde{\boldmath{\xi}}}_k }$ and $\frac{\partial \textbf{E}_{z,\Im} }{\partial \bar{\tilde{\boldmath{\xi}}}_k }$ and reducing the remaining terms yields,

\begin{eqnarray} \label{EQN:ADJOINT_DERIVATIVE_DERIVATION_STEP_3}
\frac{\mathrm{d} \tilde{\boldmath{\Phi}}}{\mathrm{d} \bar{\tilde{\boldmath{\xi}}}_k} =  \frac{\partial \textbf{E}_{z,\Re} }{\partial \bar{\tilde{\boldmath{\xi}}}_k } \left( \frac{\partial \boldmath{\Phi} }{ \partial \textbf{E}_{z,\Re} } + \boldmath{\lambda}^{\mathrm{T}} \textbf{S} +  \boldmath{\lambda}^\dagger  \textbf{S}^{*}  \right) + \frac{\partial \textbf{E}_{z,\Im} }{\partial \bar{\tilde{\boldmath{\xi}}}_k } \left( \frac{\partial \boldmath{\Phi} }{ \partial \textbf{E}_{z,\Im} } + \mathrm{i} \boldmath{\lambda}^{\mathrm{T}} \textbf{S} +  \mathrm{i} \boldmath{\lambda}^\dagger  \textbf{S}^{*}  \right)  + 2 \ \Re\left(\boldmath{\lambda}^{\mathrm{T}} \frac{\partial \textbf{S}}{\partial \bar{\tilde{\boldmath{\xi}}}_k} \textbf{E}_z \right).
\end{eqnarray}

\noindent To eliminate the first two terms in eq.~(\ref{EQN:ADJOINT_DERIVATIVE_DERIVATION_STEP_3}) containing the derivates $\frac{\partial \textbf{E}_{z,\Re} }{\partial \bar{\tilde{\boldmath{\xi}}}_k }$ and $\frac{\partial \textbf{E}_{z,\Im} }{\partial \bar{\tilde{\boldmath{\xi}}}_k }$, the two parenthesis must equal zero, 

\begin{eqnarray} \label{EQN:ADJOINT_DERIVATIVE_DERIVATION_STEP_4}
\frac{\partial \boldmath{\Phi} }{ \partial \textbf{E}_{z,\Re} } + \boldmath{\lambda}^{\mathrm{T}} \textbf{S} +  \boldmath{\lambda}^\dagger  \textbf{S}^{*} = 0, \ \ \frac{\partial \boldmath{\Phi} }{ \partial \textbf{E}_{z,\Im} } + \mathrm{i} \boldmath{\lambda}^{\mathrm{T}} \textbf{S} +  \mathrm{i} \boldmath{\lambda}^\dagger  \textbf{S}^{*} = 0,
\end{eqnarray}

\noindent multiplying the second equation by i, subtracting it from the first and transposing it yields,

\begin{eqnarray} \label{EQN:ADJOINT_DERIVATIVE_DERIVATION_STEP_5}
\frac{\partial \boldmath{\Phi} }{ \partial \textbf{E}_{z,\Re} } - \mathrm{i} \frac{\partial \boldmath{\Phi} }{ \partial \textbf{E}_{z,\Im} } + 2\boldmath{\lambda}^{\mathrm{T}} \textbf{S} = 0 \ \ \Leftrightarrow \ \ \textbf{S}^{\mathrm{T}} \boldmath{\lambda} = -\frac{1}{2} \left(  \frac{\partial \boldmath{\Phi} }{ \partial \textbf{E}_{z,\Re} } - \mathrm{i} \frac{\partial \boldmath{\Phi} }{ \partial \textbf{E}_{z,\Im} } \right)^{\mathrm{T}}.
\end{eqnarray}

\noindent Using eq.~(\ref{EQN:ADJOINT_DERIVATIVE_DERIVATION_STEP_5}) the expression in eq.~(\ref{EQN:ADJOINT_DERIVATIVE_DERIVATION_STEP_3}) reduces to the expression for $\frac{\mathrm{d} \boldmath{\Phi}}{\mathrm{d} \bar{\tilde{\boldmath{\xi}}}_k}$ in eq.~(\ref{EQN:SENSITIVITES}) and the second equation in eq.~(\ref{EQN:ADJOINT_DERIVATIVE_DERIVATION_STEP_5}) is equal to eq.~(\ref{EQN:ADJOINT_LINEAR_SYSTEM}).

\section{MATLAB code} \label{APN:MATLAB_CODE}

\begin{verbatim}
%%%%%%%%%%%%%%%%%%%%%%%%%%%%%%%%%%%%%%%%%%%%%%%%%%%%%%%%%%%%%%%%%%%%%%%%%%%
%%%%%%% A 200 LINE TOPOLOGY OPTIMIZATION CODE FOR ELECTROMAGNETISM %%%%%%%%
% --------------------------- EXAMPLE GOAL ------------------------------ %
% Designs a 2D metalens with relative permittivity eps_r capable of       %
% monocromatic focusing of TE-polarized light at a point in space.        %
% --------------------- FIGURE OF MERIT MAXIMIZED ----------------------- %
%      Phi = |Ez|^2 in a "point" (in the center of a finite element)      %
% ------------------------- EQUATION SOLVED ----------------------------- %
%      \nabla * (\nabla Ez) + k^2 A Ez = F                                %
%      With first order absorping boundary condition:                     %
%      n * \nabla Ez = - i k Ez on boundaries                             %
%      and an incident plane wave propagating from bottom to top          %
% ---------------------- DOMAIN AND DISCRETIZATION ---------------------- %
% The equation is solved in a rectangular domain, discretized using       %
% quadrilateral bi-linear finite elements                                 %
% ----------------------------------------------------------------------- %
%%%%%%%%%%%%%%%%%%%%%%%%%%%%%%%%%%%%%%%%%%%%%%%%%%%%%%%%%%%%%%%%%%%%%%%%%%%
%%%%%%%%%%%%%%%% Author: Rasmus E. Christansen, June 2020 %%%%%%%%%%%%%%%%% 
%%%%%%%%%%%%%%%%%%%%%%%%%%%%%%%%%%%%%%%%%%%%%%%%%%%%%%%%%%%%%%%%%%%%%%%%%%%
% Disclaimer:                                                             %
% The authors reserves all rights but does not guaranty that the code is  %
% free from errors. Furthermore, we shall not be liable in any event      %
% caused by the use of the program.                                       %
%%%%%%%%%%%%%%%%%%%%%%%%%%%%%%%%%%%%%%%%%%%%%%%%%%%%%%%%%%%%%%%%%%%%%%%%%%%
function [dVs,FOM]=top200EM(targetXY,dVElmIdx,nElX,nElY,dVini,eps_r,lambda,fR,maxItr)
% SETUP OF PHYSICS PARAMETERS
phy.scale = 1e-9; % Scaling finite element side length to nanometers
phy.eps_r = eps_r; % Relative permittivity
phy.k = 2*pi/(lambda*phy.scale); % Free-space wavenumber

% SETUP OF ALL INDEX SETS, ELEMENT MATRICES AND RELATED QUANTITIES
dis.nElX = nElX; % number of elements in x direction
dis.nElY = nElY; % number of elements in y direction
dis.tElmIdx = (targetXY(1)-1)*nElY+targetXY(2); % target index
dis.dVElmIdx = dVElmIdx; % design field element indices in model of physics
[dis.LEM,dis.MEM] = ELEMENT_MATRICES(phy.scale); 
[dis]=INDEX_SETS_SPARSE(dis); % Index sets for discretized model

% SETUP FILTER AND THRESHOLDING PARAMETERS
filThr.beta = 5; % Thresholding sharpness
filThr.eta = 0.5; % Thresholding level
[filThr.filKer, filThr.filSca] = DENSITY_FILTER_SETUP( fR, nElX, nElY);

% INITIALIZE DESIGN VARIABLES, BOUNDS AND OPTIMIZER OPTIONS
dVs(length(dis.dVElmIdx(:))) = dVini; % Design variables
LBdVs = zeros(length(dVs),1); % Lower bound on design variables
UBdVs = ones(length(dVs),1); % Upper bound on design variables
options = optimoptions('fmincon','Algorithm','interior-point',...
'SpecifyObjectiveGradient',true,'HessianApproximation','lbfgs',...
'Display','off','MaxIterations',maxItr,'MaxFunctionEvaluations',maxItr);

% SOLVE DESIGN PROBLEM USING MATLAB BUILT-IN OPTIMIZER: FMINCON
FOM = @(dVs)OBJECTIVE_GRAD(dVs,dis,phy,filThr);
[dVs,~] = fmincon(FOM,dVs(:),[],[],[],[],LBdVs,UBdVs,[],options);

% FINAL BINARIZED DESIGN EVALUATION
filThr.beta = 1000; 
disp('Black/white design evaluation:')
[FOM,~] = OBJECTIVE_GRAD(dVs(:),dis,phy,filThr); 
end

%%%%%%%%%%%%%%% OBJECTIVE FUNCTION AND GRADIENT EVALUATION %%%%%%%%%%%%%%%%
function [FOM,sensFOM] = OBJECTIVE_GRAD(dVs,dis,phy,filThr)
% DISTRIBUTE MATERIAL IN MODEL DOMAIN BASED ON DESIGN FIELD
dFP(1:dis.nElY,1:dis.nElX) = 0; % Design field in physics, 0: air
dFP(dis.nElY:-1:ceil(dis.nElY*9/10),1:dis.nElX) = 1; % 1: material
dFP(dis.dVElmIdx(:)) = dVs; % Design variables inserted in design field

% FILTERING THE DESIGN FIELD AND COMPUTE THE MATERIAL FIELD
dFPS = DENSITY_FILTER(filThr.filKer,filThr.filSca,dFP,ones(dis.nElY,dis.nElX));
dFPST = THRESHOLD( dFPS, filThr.beta, filThr.eta);
[A,dAdx] = MATERIAL_INTERPOLATION(phy.eps_r,dFPST,1.0); % Material field

% CONSTRUCT THE SYSTEM MATRIX
[dis,F] = BOUNDARY_CONDITIONS_RHS(phy.k,dis,phy.scale);
dis.vS = reshape(dis.LEM(:)-phy.k^2*dis.MEM(:)*(A(:).'),16*dis.nElX*dis.nElY,1);
S = sparse([dis.iS(:);dis.iBC(:)],[dis.jS(:);dis.jBC(:)],[dis.vS(:);dis.vBC(:)]);

tic;
% SOLVING THE STATE SYSTEM: S * Ez = F
[L,U,Q1,Q2] = lu(S); % LU - factorization
Ez = Q2 * (U\(L\(Q1 * F)));  Ez = full(Ez); % Solving
toc;

% FIGURE OF MERIT 
P = sparse(dis.edofMat(dis.tElmIdx,:),dis.edofMat(dis.tElmIdx,:),1/4,...
(dis.nElX+1)*(dis.nElY+1),(dis.nElX+1)*(dis.nElY+1)); % Weighting matrix
FOM = Ez' * P * Ez; % Solution in target element

% ADJOINT RIGHT HAND SIDE
AdjRHS = P*(2*real(Ez) - 1i*2*imag(Ez));

% SOLVING THE ADJOING SYSTEM: S.' * AdjLambda = AdjRHS
AdjLambda = (Q1.') * ((L.')\((U.')\((Q2.') * (-1/2*AdjRHS)))); % Solving

% COMPUTING SENSITIVITIES
dis.vDS = reshape(-phy.k^2*dis.MEM(:)*(dAdx(:).'),16*dis.nElX*dis.nElY,1); 
DSdx = sparse(dis.iElFull(:),dis.jElFull(:),dis.vDS(:)); % Constructing dS/dx
DSdxMulV = DSdx * Ez(dis.idxDSdx); % Computing dS/dx * Field values
DsdxMulV = sparse(dis.iElSens,dis.jElSens,DSdxMulV);
sens = 2*real(AdjLambda(dis.idxDSdx).' * DsdxMulV); % Computing sensitivites
sens = full(reshape(sens,dis.nElY,dis.nElX));

% FILTERING SENSITIVITIES
DdFSTDFS = DERIVATIVE_OF_THRESHOLD( dFPS, filThr.beta, filThr.eta);
sensFOM = DENSITY_FILTER(filThr.filKer,filThr.filSca,sens,DdFSTDFS);

% EXTRACTING SENSITIVITIES FOR DESIGNABLE REGION
sensFOM = sensFOM(dis.dVElmIdx);

% FMINCON DOES MINIMIZATION
FOM = -FOM; sensFOM = -sensFOM(:);

% PLOTTING AND PRINTING
figure(1); % Field intensity, |Ez|^2
imagesc((reshape(Ez.*conj(Ez),dis.nElY+1,dis.nElX+1))); colorbar; axis equal;
figure(2); % Physical design field
imagesc(1-dFPST); colormap(gray); axis equal; drawnow;
disp(['FOM: ' num2str(-FOM)]); % Display FOM value
end

%%%%%%%%%%%%%%%%%%%%%%%%%% AUXILIARY FUNCTIONS %%%%%%%%%%%%%%%%%%%%%%%%%%%%
%%%%%%%%%%%% ABSORBING BOUNDARY CONDITIONS AND RIGHT HAND SIDE %%%%%%%%%%%%
function [dis,F] = BOUNDARY_CONDITIONS_RHS(waveVector,dis,scaling)
AbsBCMatEdgeValues = 1i*waveVector*scaling*[1/6 ; 1/6 ; 1/3 ; 1/3];
% ALL BOUNDARIES HAVE ABSORBING BOUNDARY CONDITIONS
dis.iBC = [dis.iB1(:);dis.iB2(:);dis.iB3(:);dis.iB4(:)];
dis.jBC = [dis.jB1(:);dis.jB2(:);dis.jB3(:);dis.jB4(:)];
dis.vBC = repmat(AbsBCMatEdgeValues,2*(dis.nElX+dis.nElY),1);
% BOTTOM BOUNDARY HAS INCIDENT PLANE WAVE
F = zeros((dis.nElX+1)*(dis.nElY+1),1); % System right hand side
F(dis.iRHS(1,:)) = F(dis.iRHS(1,:))+1i*waveVector;
F(dis.iRHS(2,:)) = F(dis.iRHS(2,:))+1i*waveVector;
F = scaling*F;
end
%%%%%%%%%%%%%%%%%%%%% CONNECTIVITY AND INDEX SETS %%%%%%%%%%%%%%%%%%%%%%%%%
function [dis]=INDEX_SETS_SPARSE(dis)
% INDEX SETS FOR SYSTEM MATRIX 
nEX = dis.nElX; nEY = dis.nElY; % Extracting number of elements
nodenrs = reshape(1:(1+nEX)*(1+nEY),1+nEY,1+nEX); % Node numbering
edofVec = reshape(nodenrs(1:end-1,1:end-1)+1,nEX*nEY,1); % First DOF in element
dis.edofMat = repmat(edofVec,1,4)+repmat([0 nEY+[1 0] -1],nEX*nEY,1);
dis.iS = reshape(kron(dis.edofMat,ones(4,1))',16*nEX*nEY,1);
dis.jS = reshape(kron(dis.edofMat,ones(1,4))',16*nEX*nEY,1);
dis.idxDSdx = reshape(dis.edofMat',1,4*nEX*nEY);
% INDEX SETS FOR BOUNDARY CONDITIONS 
TMP = repmat([[1:nEY];[2:nEY+1]],2,1);
dis.iB1 = reshape(TMP,4*nEY,1); % Row indices
dis.jB1 = reshape([TMP(2,:);TMP(1,:);TMP(3,:);TMP(4,:)],4*nEY,1); % Column indices
TMP = repmat([1:(nEY+1):(nEY+1)*nEX;(nEY+1)+1:(nEY+1):(nEY+1)*nEX+1],2,1);
dis.iB2 = reshape(TMP,4*nEX,1);
dis.jB2 = reshape([TMP(2,:);TMP(1,:);TMP(3,:);TMP(4,:)],4*nEX,1);
TMP = repmat([(nEY+1)*(nEX)+1:(nEY+1)*(nEX+1)-1;(nEY+1)*(nEX)+2:(nEY+1)*(nEX+1)],2,1);
dis.iB3 = reshape(TMP,4*nEY,1);
dis.jB3 = reshape([TMP(2,:);TMP(1,:);TMP(3,:);TMP(4,:)],4*nEY,1);
TMP = repmat([2*(nEY+1):nEY+1:(nEY+1)*(nEX+1);(nEY+1):nEY+1:(nEY+1)*(nEX)],2,1);
dis.iB4 = reshape(TMP,4*nEX,1);
dis.jB4 = reshape([TMP(2,:);TMP(1,:);TMP(3,:);TMP(4,:)],4*nEX,1);
dis.iRHS = TMP;
% INDEX SETS FOR INTEGRATION OF ALL ELEMENTS
ima0 = repmat([1,2,3,4,1,2,3,4,1,2,3,4,1,2,3,4],1,nEX*nEY).';
jma0 = repmat([1,1,1,1,2,2,2,2,3,3,3,3,4,4,4,4],1,nEX*nEY).';
addTMP = repmat(4*[0:nEX*nEY-1],16,1);
addTMP = addTMP(:);
dis.iElFull = ima0+addTMP;
dis.jElFull = jma0+addTMP;
% INDEX SETS FOR SENSITIVITY COMPUTATIONS
dis.iElSens = [1:4*nEX*nEY]';
jElSens = repmat([1:nEX*nEY],4,1);
dis.jElSens = jElSens(:);
end
%%%%%%%%%%%%%%%%%% MATERIAL PARAMETER INTERPOLATION %%%%%%%%%%%%%%%%%%%%%%%
function [A,dAdx] = MATERIAL_INTERPOLATION(eps_r,x,alpha_i)
A = 1 + x*(eps_r-1) - 1i * alpha_i * x .* (1 - x); % Interpolation
dAdx = (eps_r-1)*(1+0*x) - 1i * alpha_i * (1 - 2*x); % Derivative of interpolation
end
%%%%%%%%%%%%%%%%%%%%%%%%%%% DENSITY FILTER  %%%%%%%%%%%%%%%%%%%%%%%%%%%%%%%
function [xS]=DENSITY_FILTER(FilterKernel,FilterScaling,x,func)
xS = conv2((x .* func)./FilterScaling,FilterKernel,'same');
end
function [ Kernel, Scaling ] = DENSITY_FILTER_SETUP( fR, nElX, nElY )
[dy,dx] = meshgrid(-ceil(fR)+1:ceil(fR)-1,-ceil(fR)+1:ceil(fR)-1);
Kernel = max(0,fR-sqrt(dx.^2+dy.^2)); % Cone filter kernel
Scaling = conv2(ones(nElY,nElX),Kernel,'same'); % Filter scaling
end
%%%%%%%%%%%%%%%%%%%%%%%%%%%% THRESHOLDING %%%%%%%%%%%%%%%%%%%%%%%%%%%%%%%%%
function [ xOut ] = THRESHOLD( xIn, beta, eta)
xOut = (tanh(beta*eta)+tanh(beta*(xIn-eta)))./(tanh(beta*eta)+tanh(beta*(1-eta)));
end
function [ xOut ] = DERIVATIVE_OF_THRESHOLD( xIn, beta, eta)
xOut = (1-tanh(beta*(xIn-eta)).^2)*beta./(tanh(beta*eta)+tanh(beta*(1-eta)));
end
%%%%%%%%%%%%%%%%%%%%%%%%%%% ELEMENT MATRICES %%%%%%%%%%%%%%%%%%%%%%%%%%%%%%
function [LaplaceElementMatrix,MassElementMatrix] = ELEMENT_MATRICES(scaling)
% FIRST ORDER QUADRILATERAL FINITE ELEMENTS
aa=scaling/2; bb=scaling/2; % Element size scaling
k1=(aa^2+bb^2)/(aa*bb); k2=(aa^2-2*bb^2)/(aa*bb); k3=(bb^2-2*aa^2)/(aa*bb);
LaplaceElementMatrix = [k1/3 k2/6 -k1/6 k3/6 ; k2/6 k1/3 k3/6 -k1/6; ...
-k1/6  k3/6  k1/3  k2/6; k3/6 -k1/6  k2/6  k1/3];
MassElementMatrix = aa*bb*[4/9 2/9 1/9 2/9 ; 2/9 4/9 2/9 1/9 ; ...
1/9 2/9 4/9 2/9; 2/9 1/9 2/9 4/9];
end
\end{verbatim}

\section*{Funding}

This work was supported in by Villum Fonden through the NATEC (NAnophotonics for TErabit Communications) Centre (grant no.~8692) and by the Danish National Research Foundation through NanoPhoton Center for Nanophotonics (grant no.~DNRF147).

\section*{Disclosures} 

The authors declare that there are no conflicts of interest related to this article.

\bibliographystyle{ieeetr} 
\bibliography{References}

\end{document}